\documentclass[useAMS,usenatbib]{mn2e}
\usepackage{epsfig}
\usepackage{times}
\usepackage{longtable,lscape}
\usepackage{color}

\def\cm3{cm$^{-3}$}
\def\kms{km~s$^{-1}$}
\def\msunyr{M$_{\odot}$\,yr$^{-1}$}

\def\msun{M$_{\odot}$}

\def\beq{\begin{equation}}
\def\eeq{\end{equation}}

\def\lesssim{\mathrel{\hbox{\rlap{\hbox{\lower4pt\hbox{$\sim$}}}\hbox{$<$}}}}
\def\gtrsim{\mathrel{\hbox{\rlap{\hbox{\lower4pt\hbox{$\sim$}}}\hbox{$>$}}}}

\def\aj{AJ}
\def\pasp{PASP}
\def\apj{ApJ}
\def\apjs{ApJS}
\def\apjl{ApJL}
\def\aap{A\&A}
\def\araa{ARA\&A}
\def\aaps{A\&AS}
\def\mnras{MNRAS}

\def\apss{Ap\&SS}

\definecolor{red}{rgb}{1,0,0}

\title[]{Determining the main-sequence mass of Type II supernova progenitors}

\author[Luc Dessart, Eli Livne, and Roni Waldman]
{
Luc Dessart$^{1}$\thanks{E-mail: Luc.Dessart@oamp.fr},
Eli Livne$^{2}$, and Roni Waldman$^{2}$\\
$^{1}$ Laboratoire d’Astrophysique de Marseille, Universit\'e de Provence,
CNRS, 38 rue Fr\'ed\'eric Joliot-Curie, F-13388 Marseille Cedex 13, France \\
$^{2}$ Racah Institute of Physics, The Hebrew University, Jerusalem, Israel
}

\begin{document}

\date{Accepted . Received }

\pagerange{\pageref{firstpage}--\pageref{lastpage}} \pubyear{2010}

\maketitle

\label{firstpage}

\begin{abstract}

We present radiation-hydrodynamics simulations of core-collapse supernova
(SN) explosions, artificially generated by driving a piston at the base of the envelope
of a rotating or non-rotating red-supergiant progenitor
star. We search for trends in ejecta kinematics in the resulting Type
II-Plateau (II-P) SN, exploring dependencies with explosion energy and pre-SN
stellar-evolution model.

We recover the trivial result that larger explosion energies yield larger
ejecta velocities in a given progenitor. However, we emphasise that
for a {\it given} explosion energy, the increasing helium-core mass
with main-sequence mass of such Type II-P SN progenitors leads to ejection
of core-embedded oxygen-rich material at larger velocities.
We find that the photospheric velocity at 15\,d after shock breakout is a good
and simple indicator of the explosion energy in our selected set of pre-SN
models. This measurement, combined with the width of the nebular-phase O{\sc i}\,6303--6363\AA\ line,
can be used to place an upper-limit on the progenitor main-sequence mass.
Using the results from our simulations, we find that the current, but remarkably scant, late-time spectra
of Type II-P SNe support progenitor main-sequence masses inferior
to $\sim$20\,\msun, and thus, corroborate the inferences based on the direct, but difficult, progenitor
identification in pre-explosion images. The narrow width of  O{\sc i}\,6303--6363\AA\
in Type II-P SNe with nebular spectra does not support high-mass progenitors in the range 25--30\,\msun.

Combined with quantitative spectroscopic modelling, such diagnostics offer a
means to constrain the main-sequence mass of the progenitor, the mass
fraction of the core ejected, and thus, the mass of the compact remnant formed.
\end{abstract}

\begin{keywords} radiation hydrodynamics -- stars: atmospheres -- stars:
supernovae - stars: transients
\end{keywords}

\section{Introduction}

   Understanding how and which massive stars end their lives in a successful core-collapse SN explosion
is still subject to large uncertainties today. Various mechanisms have been proposed, all invoking
the gravitational, and sometimes rotational, energy of the collapsed core and infalling mantle. The hot
proto-neutron star (PNS) that forms at core bounce releases on a timescale of $\sim$10\,s
on the order of 100\,B (1\,B$\equiv$10$^{51}$\,erg) of energy in the form of neutrinos. 
A neutrino-driven explosion hinges on the successful absorption of a few percent of this neutrino energy 
behind the shock (see, e.g., \citealt{herant_etal_94}). This must occur prior to black-hole formation,
and thus within a few seconds at most of core bounce.
This does not seem to occur {\it robustly} in any of the one- or two-dimensional radiation-hydrodynamics
simulations performed so far with sophisticated neutrino transport and microphysics,
although the latest simulations look promising \citep{marek_janka_09}.
\citet[but see \citealt{weinberg_quataert_08} and \citealt{marek_janka_09}]{burrows_etal_06}
proposed that if such a neutrino-driven explosion fails within a second of core bounce,
the stalled shock may instead be energised by the acoustic power associated with gravity-mode oscillations
of the PNS. In such simulations, core oscillations could power an explosion over a wide range of
massive-star progenitors, as obtained for objects with a main-sequence mass between 11 and 25\,\msun\ by
\citet{burrows_etal_07a}.
Finally, provided the progenitor star possesses a very fast-rotating core at the onset of collapse
(typically with a 1-s rotation period), a magneto-rotational explosion may arise within a few hundred milliseconds of
core bounce, with a potential to produce an axisymmetric and highly energetic explosion
\citep{leblanc_wilson_70,BPS_76,symbalisty_84,akiyama_etal_03,moiseenko_etal_06,uzdenski_mcfadyen_07,
burrows_etal_07b,dessart_etal_08b}. A variety of potential mechanisms thus exists to power a wide range of
core-collapse SN explosions, with energy perhaps as low as 0.01\,B and as high as 10\,B,
and with a morphology departing quite naturally from spherical symmetry. However, it is currently
difficult to go beyond such general statements, by quantitatively
characterising the specific properties of the progenitor stars associated with the core-collapse SNe
we see and by assessing how they exploded.

An important question is to determine if there is an upper-limit to the main-sequence star mass for a successful core-collapse
SN explosion. This is relevant for determining the chemical yields and the
properties/identity of the compact object produced by the event, but also
for the understanding of the mechanism of explosion and its dependency on, e.g., stellar structure.
Currently, identification of the SN progenitor on pre-explosion images has been
possible for a handful of cases, and primarily for the Type II-Plateau (II-P) SNe,
confirming their progenitors are red-supergiant (RSG) stars (see, e.g., \citealt{vandyck_etal_03a,smartt_etal_04,maund_etal_05,
li_etal_06,mattila_etal_08}).
Surprisingly, such observations also suggest that these progenitors have a main-sequence mass
in the range $\sim$8.5-16.5\,\msun\ (see \citealt{smartt_09} and references therein), which is at the
lower end of the range expected from stellar evolutionary calculations.
On the other hand, radiation-hydrodynamics simulations of SNe II-P light curves and their photospheric-velocity
evolution generally support larger masses, at odds with observational inferences (see, e.g., \citealt{utrobin_chugai_09}).
An issue here is our ability to distinguish between inferences on the final mass (that at core collapse), and the initial
mass (that on the main sequence).

  Observationally, and until neutrino and gravitational-wave detectors obtain their novel signatures associated
with a core-collapse SN, photons represent the main source of information for inferring the properties of the ejecta and
their progenitors. Because of its relative simplicity, bolometric light-curve modelling has been the main tool.
In the context of hydrogen-rich core-collapse SNe, the emergent radiation is primarily conditioned
 by the properties of the progenitor star, i.e. its final mass and in particular the mass/size of its hydrogen-envelope,
 and the properties of the explosion, i.e. the energetics and the mass of nucleosynthesized $^{56}$Ni
 \citep{falk_arnett_77,popov_93,nadyozhin_03,baklanov_etal_05,utrobin_07,
 kasen_woosley_09,DLW_10}. These observables appear, however, quite degenerate
 (see, e.g., \citealt{hamuy_03,bersten_hamuy_09}) and do not seem to manifest much dependency
 on the properties of the progenitor helium core. Their interpretation is
  complicated by multiple factors associated with the explosion mechanism, as discussed above,
  but also by the uncertainties in the pre-SN star mass, which is affected by mass loss.

  Stellar-wind mass loss, in particular during the RSG phase, peels
  off the hydrogen envelope of a 10--30\,\msun\ progenitor massive star and leaves it with a much reduced mass
  (this reduction depends on stellar mass) by the time of core collapse.
  The mean RSG mass-loss-rate value appears to be on the order of 10$^{-5}$--10$^{-6}$\,\msunyr\ \citep{dejager_etal_88},
  but large uncertainties exist owing to the incomplete knowledge of the mass loss mechanism (see, e.g., \citealt{josselin_plez_07}).
   The cumulative loss of mass is conditioned by the duration of that RSG phase, which can be lengthened
   by stellar rotation \citep{meynet_etal_06}.
    These objects may also go through ``super-wind" phases with mass-loss rates as high as $\sim$10$^{-4}$\,\msunyr,
    in particular in the ultimate stages of evolution \citep{heger_etal_98,yoon_cantiello_10}.
    The presence of massive nebulae around such RSG stars
    suggests that dramatic mass-loss events can take place \citep{smith_etal_09}. In some, they may even be triggered
    by nuclear flashes at the surface of the degenerate core \citep{weaver_woosley_79}, with dramatic consequences
    for their loosely-bound massive hydrogen-rich envelopes \citep{DLW_10}.
    When extended to include binary-star evolution channel, these issues become even more complicated.
    The likely important role of binary-star evolution to the making of SN progenitors remains to be thoroughly explored
    (in the context of SNe associated with a $\gamma$-ray burst, see, e.g., \citealt{cantiello_etal_07}).
    Overall, these complications make the final mass and radius of the pre-SN II-P progenitor uncertain.

     In contrast, a key property of single-massive stars is their increasing helium-core mass with main-sequence
     mass (and rotation).
     When considering SN II progenitors,
     the helium-core mass is not affected by mass loss since a residual hydrogen envelope
     is still present. In Type Ib/c SN progenitors, mass loss is thought to have not
     only peeled away the hydrogen envelope, but also the helium core itself, braking the unique
     association between progenitor mass and helium-core mass.\footnote{Binary-star
     evolution, with mass transfer and the possible evolution through a
     common envelope phase, is a further complication \citep{taam_ricker_06}. However, the final product might
     not die as a RSG and not give rise to a SN II-P.}
     Hence, in SNe II, while there may be doubt on the mass of hydrogen
     endowed by a given massive star at collapse (and more generally the final mass),
     the mass of the helium core is very strictly set by
     stellar structure/evolution and is independent of the mass-loss history of the star.
     In the range between 11 and 30\,\msun, this helium-core mass grows enormously, from
     around 1.5--2\,\msun\ up to 8-10\,\msun\ \citep{woosley_weaver_95,WHW02,arnett_91}.
     Hence, ignoring for now stellar rotation, the main-sequence mass ties directly with the helium-core mass, and could be inferred
     provided the latter is somehow constrained from the SN light.

\begin{table*}
\begin{minipage}{170mm}
\caption{Summary of the properties for a representative sample of pre-SN models employed in this study.
From the upper part to the lower part of the table, we show properties for non-rotating (``s" series; WHW02)
and rotating (``E" series; HLW00) pre-SN models.
For each, we give the initial mass $M_{\rm i}$ (i.e. the main-sequence mass),
the final mass $M_{\rm f}$ (i.e. the mass at core collapse), the Lagrangian mass delimiting important shells
(outer edge of the core $M_{\rm core}$; outer edge of the oxygen shell $M_{\rm e,O}$;
outer edge of the helium shell $M_{\rm e,He}$;  inner edge of the hydrogen shell $M_{\rm i,H}$),
the mass $M$(H env.)  and the size $\Delta R$(H env.) of the hydrogen-rich envelope,
the surface radius $R_{\star}$, the gravitational and binding energies ($E_{\rm grav}$ and $E_{\rm binding}$)
of the progenitor envelope outside of $M_{\rm core}$.
[See text for discussion] \label{tab_sum_progenitor}}
\begin{tabular}{ccccccccccccc}
\hline
Pre-SN Model & $M_{\rm i}$ & $M_{\rm f}$  & $M_{\rm core}$ & $M_{\rm e,O}$ & $M_{\rm e,He}$  & $M_{\rm i,H}$ & $M$(H env.)  & $\Delta R$(H env.) & $R_{\star}$ & $E_{\rm grav}$  & $E_{\rm binding}$ \\
                           &  [\msun]        &  [\msun]        &            [\msun]        &       [\msun]        &          [\msun]         &      [\msun]        &      [\msun]        & [10$^{13}$\,cm]& [10$^{13}$\,cm] & [B]                       &      [B] \\
\hline
s11 & 11.0 &   10.61  &    1.37  &    1.69  &    1.75  &    2.78  &    7.83  &    4.06  &   4.086  &   0.342  &   0.135 \\
s15 & 15.0 &   12.64  &    1.62  &    2.61  &    3.00  &    4.15  &    8.49  &    5.86  &   5.867  &   0.995  &   0.392 \\
s20 & 20.0 &   14.73  &    1.46  &    4.25  &    4.95  &    6.14  &    8.59  &    7.81  &   7.816  &   1.411  &   0.399 \\
s25 & 25.0 &   12.53  &    1.92  &    5.71  &    7.15  &    8.40  &    4.13  &   10.06  &  10.066  &   3.310  &   0.858 \\
s30 & 30.0 &    12.24  &    2.04  &    7.72  &    9.18  &   10.02  &    2.22  &    8.51  &   8.516  &   2.984  &   0.744 \\
\hline
E10 & 10.0 & 9.23  &    1.48  &    1.72  &    1.82  &    2.89  &    6.34  &    3.82  &   3.842  &   0.354  &   0.170 \\
E12 & 12.0 & 10.35  &    1.50  &    2.31  &    3.49  &    3.72  &    6.63  &    5.26  &   5.271  &   0.637  &   0.213 \\
E15 & 15.0 & 10.86  &    1.62  &    3.39  &    4.94  &    5.19  &    5.68  &    7.51  &   7.515  &   1.318  &   0.402 \\
E20 & 20.0 & 11.01  &    1.84  &    4.38  &    7.75  &    7.96  &    3.05  &   15.63  &  15.643  &   2.465  &   0.718 \\
\hline
\end{tabular}
\end{minipage}
\end{table*}

    The amount of information that can be extracted from SN II-P light curves alone is limited.
    This stems from the fact that the {\it plateau phase} that is generally the focus of attention
    provides restricted information, primarily on the shocked hydrogen-envelope, whose pre-shocked properties
    are made uncertain because of mass loss.
    It provides no unambiguous information on chemical yields,
    no information on the distribution of elements in velocity space or on the properties of the progenitor helium core,
    no direct evidence
    for mixing etc. In contrast, having recourse to both SN spectra and light-curves, together with sophisticated
    radiative-transfer and radiation-hydrodynamics tools to model them, can alleviate these shortcomings.
    This modelling approach can apply even for distant Type II-P SNe and is thus advantageous over the
    direct identification of progenitors on pre-explosion images, which only works for nearby objects
    (perhaps up to $\sim$20\,Mpc).

    For this study, we generated a wide range of SN II-P explosions, using the one-dimensional grey radiation-hydrodynamics
    code {\sc v1d} \citep{livne_93,DLW_10}, and starting our simulations from the pre-SN models for non-rotating and rotating
    massive stars computed {\it at solar metallicity}
    by \citet[hereafter, WHW02]{WHW02} and \citet[hereafter, HLW00]{heger_etal_00}, respectively.
    Leaving to a forthcoming paper the full presentation of non-local thermodynamic equilibrium (non-LTE) 
    synthetic spectra and light curves (Dessart et al,
    in preparation), we focus the discussion here on the properties of such SN II-P ejecta, in particular
    the chemical distribution in velocity space. This has already been discussed in the past, but usually
    merely ``in passing", by studies of chemical yields of core-collapse SN explosions in the modern or early Universe
    \citep{woosley_weaver_95,tominaga_etal_07,heger_woosley_08,tominaga_09,joggerst_etal_10},
    chemical mixing by Rayleigh-Taylor instabilities in the first few hours of the life of the SN \citep{herant_woosley_94,mueller_etal_91,
    kifonidis_etal_00,kifonidis_etal_03,kifonidis_etal_06,hammer_etal_09}, or in the context of the remnant mass (see, e.g. \citealt{zhang_etal_08}).
    Here, we emphasise that strong constraints can be placed on the main-sequence mass of a SN II-P progenitor
    from the ejection-speed of core-embedded oxygen-rich material and the expansion-rate of the hydrogen-rich progenitor envelope.

    This paper is structured as follows. In \S\ref{sect_model}, we summarise the properties of the progenitor models
    used in this study, as well as the numerical setup for our
    one-dimensional grey radiation-hydrodynamics simulations using the code {\sc v1d} \citep{livne_93,DLW_10}.
    We then present in \S\ref{sect_res} the ejecta kinematics we obtain for a variety of progenitor stars and explosion parameters,
    and discuss how such properties can be used to constrain the progenitor main-sequence mass.
    Separately, in an  appendix, we present other results obtained from this grid of models.
    In \S\ref{sect_concl}, we conclude and confront our results to observations. In particular, we discuss how one can attempt
    to spectroscopically constrain a SN II-P progenitor main-sequence mass from the observation of nebular-phase
    O{\sc i}\,6303-6363\AA\ and photospheric-phase optical spectra.

\section{Progenitor models and Numerical Setup}
\label{sect_model}

\subsection{Progenitor models}

    Our radiation-hydrodynamics study employs as starting conditions two different sets of stellar evolutionary calculations
for massive stars at solar metallicity and with allowance for mass loss.
We use both the non-rotating pre-SN models of WHW02, with main-sequence
mass in the range 11--30\,\msun\ (model prefix ``s"), and the rotating pre-SN models of HLW00, with main-sequence mass
in the range 10--20\,\msun\ (model prefix ``E"; the initial equatorial velocity is 200\,\kms).
We refer the reader to the corresponding references for details on these evolutionary calculations.
In each case, the evolution is computed until the onset of collapse of the degenerate core.
This choice of mass range is motivated by our desire to select only those progenitors that have retained
a hydrogen envelope massive enough to produce a SN II-P when exploded.
Hence, for all of these pre-SN models, mass loss has not had any erosive effect on the helium-core,
whose mass is thus an increasing function of main-sequence mass.

\begin{figure*}
\epsfig{file=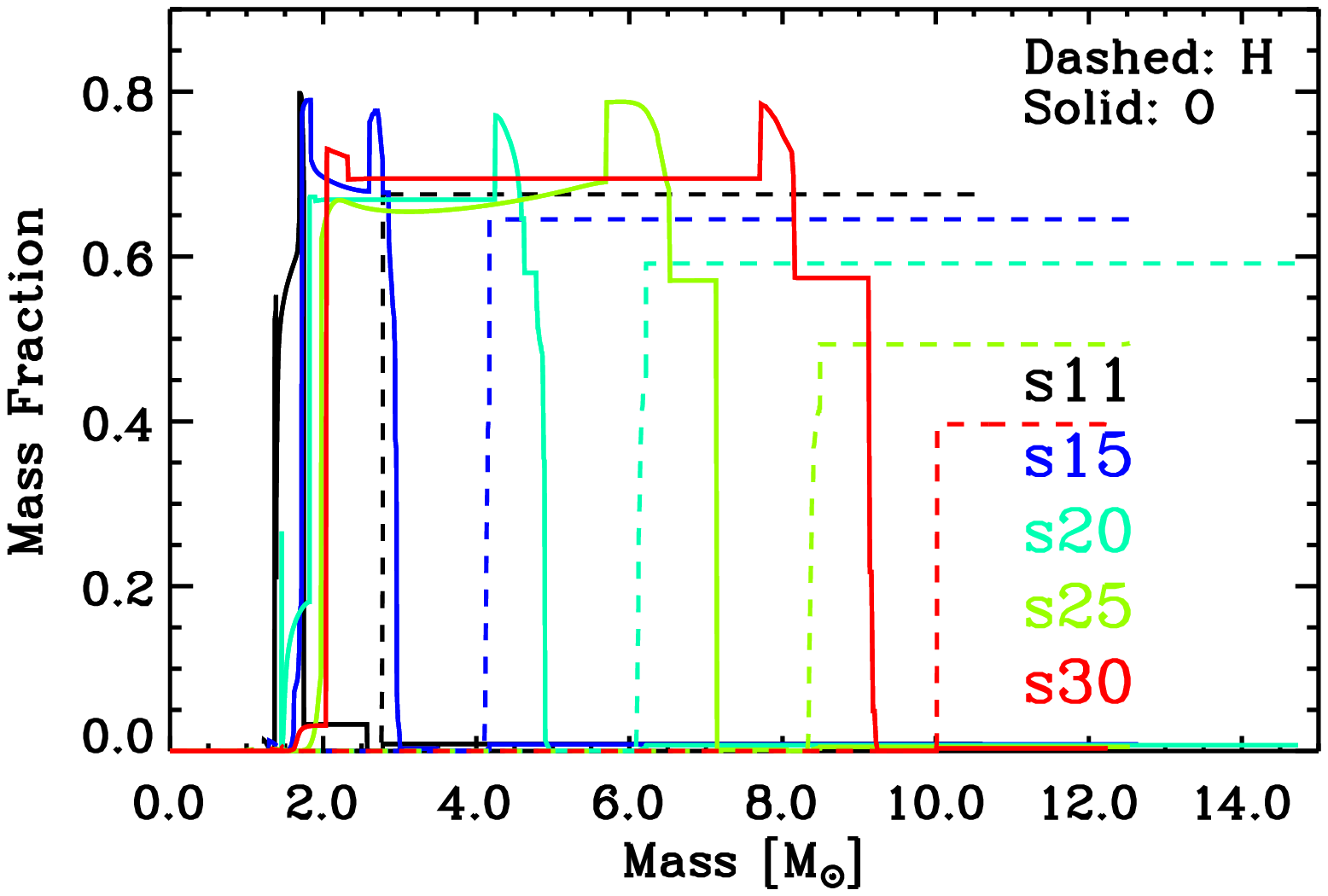,width=8.5cm}
\epsfig{file=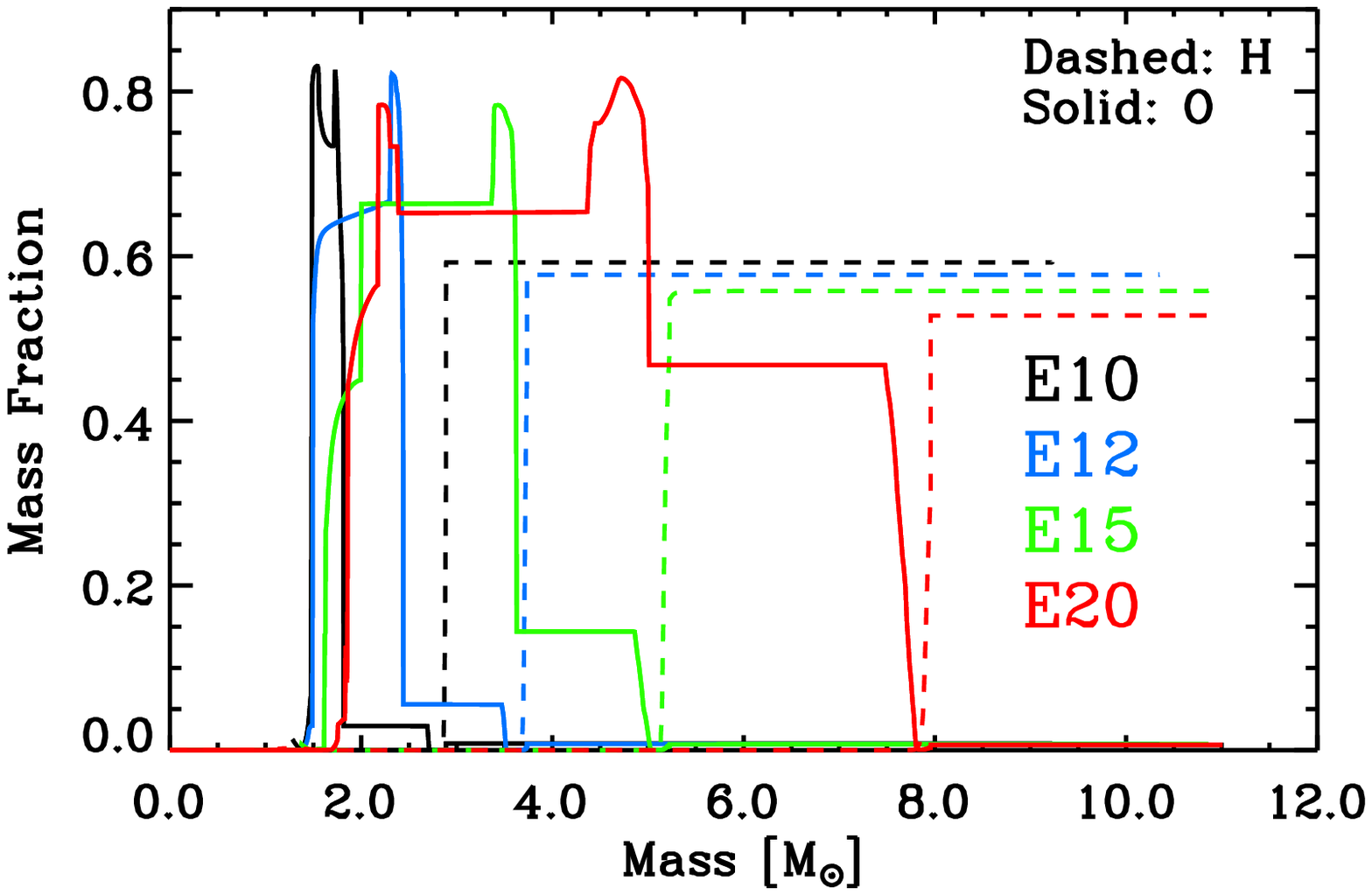,width=8.5cm}
\caption{Illustration of the distribution of hydrogen (dashed) and oxygen (solid) versus Lagrangian mass in the
envelope of a few non-rotating (left; ``s" series from WHW02) and rotating (right; ``E" series from HLW00) pre-SN
models. A colour coding is used to differentiate models of differing main-sequence mass
(given by the number XX in sXX or EXX; see Table~\ref{tab_sum_progenitor}). Notice the growing
mass of the oxygen-rich shell with increasing main-sequence mass, and for a given main-sequence mass when
rotation is included. [See text for discussion]. \label{fig_chem_strat}}
\end{figure*}

In Table~\ref{tab_sum_progenitor}, we give a summary of the properties for a
representative sample of the (non-rotating) WHW02 pre-SN models,
i.e. models s11, s15, s20, s25, and s30, as well as those for our set of (rotating) HLW00 pre-SN models (E10, E12, E15, and E20).
We give the approximate edge location of the oxygen/helium/hydrogen shells, adopting the depth where the mean molecular weight
drops outward below 16.5/4.5/2.0, respectively.
Note in particular that, for non-rotating models in the mass range 11--30\,\msun, the helium core mass
grows from 1.75 up to 9.18\,\msun, while the final star mass at collapse varies little and is in the range
10.61--14.73\,\msun.
Hence, in these non-rotating models,  the helium-core mass represents from 16\% up to 75\%
of the total star mass at the onset of collapse (the relative fraction of the star that represents the
hydrogen-rich envelope decreases correspondingly).
We illustrate the chemical stratification of these pre-SN models in the left panel of Fig.~\ref{fig_chem_strat},
selecting oxygen and hydrogen for better visibility.
For more massive main-sequence objects, oxygen is globally more abundant and is present closer (in mass coordinate)
to the surface.

In the lower part of Table~\ref{tab_sum_progenitor}, we summarise the properties of the rotating pre-SN models of HLW00,
showing the envelope chemical stratification in the right panel of Fig.~\ref{fig_chem_strat}.
Rotation produces bigger helium cores for a given star mass, so that models E15 and s20 have similar helium-core masses
despite having 15 and 20\,\msun\ main-sequence masses, respectively.
Furthermore, because of enhanced mixing and mass loss, the maximum mass for a rotating star to produce a SN II-P
is lowered. Here, the HLW00 models suggest that stars initially more massive than $\sim$20\,\msun\ will not retain
any hydrogen by the time of collapse and thus cannot lead to a SN II event.

A generic feature of all these hydrogen-rich massive stars is that their hydrogen-rich envelope
is always very loosely bound \citep{DLW_10}. It
is very extended and represents typically 99\% of the size of the star, so that the helium core is
contained in the inner few percent of the stellar radius.
The gravitational energy scaling with the square of the mass (but only with the inverse of the radius),
the helium core becomes increasingly more bound in higher-mass progenitors.
With the increase of the helium core mass with main-sequence mass, the mean envelope binding energy
of the pre-SN progenitor star increases with main-sequence mass (see Table~\ref{tab_sum_progenitor}).
Ultimately, this conditions the successful ejection of  helium-core material and the magnitude of fallback
\citep{woosley_weaver_95,zhang_etal_08}.

\subsection{Radiation-hydrodynamics simulations with {\sc v1d}}

Starting with the pre-SN progenitor models described above, we then employ
the one-dimensional grey (i.e. one-group) flux-limited-diffusion radiation-hydrodynamics code {\sc v1d}
to simulate a core-collapse SN explosion.
A description of this version of the code is given in \citet{DLW_10} and will not be repeated here since the
numerical approach is basically the same, apart from the energy deposition procedure.
In the past, our approach was to simply deposit internal energy at a specified rate and up to a given
amount at the base of the grid, chosen at some mass cut. For this study, we find it
more practical to generate the explosion by driving a piston at the grid base. This is the
standard, albeit artificial, way of generating a SN ejecta when modeling SN light curves and
asymptotic ejecta properties. Excising the core region alleviates the Courant-time limitation and permits
the simulation of one model out to one year within 1--2 days on a single processor.
Following the collapse, bounce, and post-bounce phases of the entire object all the way to one year
would instead be a computational challenge, also compromised by the incomplete knowledge
of the explosion mechanism.

To set the piston properties (mass cut $M_{\rm piston}$ and speed $V_{\rm piston}$),
we guide ourselves with results from multi-dimensional (neutrino)
radiation-hydrodynamics  simulations of core-collapse SN explosions.
Whatever the explosion mechanism (see introduction
for a concise summary), recent simulations suggest the explosion appears with a delay
of a few hundred milliseconds after core bounce. This delay is necessary to increase the neutrino-energy deposition in the gain region
for neutrino-driven explosions (see, e.g., \citealt{marek_janka_09}), or to achieve the necessary magnetic-field
amplification in magneto-rotational explosions of fast-rotating progenitor stars
(see, e.g., \citealt{burrows_etal_07b}). Owing to the different iron-core structure of massive stars of various main-sequence mass,
this will correspond to different mass cuts at the onset of explosion.
In practice, we determine the piston mass cut by performing core-collapse simulations for each 
of our WHW02/HLW00 pre-SN model
using the code {\sc gr1d} \citep{gr1d}. The simulations were done using the stiff nuclear Equation of State
they provide, general-relativistic gravity, and a leakage scheme (combined with a parameterisation of
the electron fraction with mass density) for the treatment of deleptonisation.
In all such simulations, the PNS mass rapidly grows after core bounce
due to the huge initial accretion rates, but this mass evolves slowly past a few hundred milliseconds post-bounce time.
Furthermore, at such times, the enclosed mass between the PNS surface and a radius of 1000-2000\,km is a small
fraction of the PNS mass. Hence, we assigned to $M_{\rm piston}$ the
value obtained in our {\sc gr1d} simulations for the accumulated baryonic mass enclosed in the inner 500\,km at
a post-bounce time of 500\,ms (fourth-column entry in Tables~2--6). Our resulting choice for $M_{\rm piston}$ agrees  to within
0.1--0.2\,\msun\ with the values used by \citet{woosley_weaver_95} or more recently by \citet{zhang_etal_08}.
We find that variations of a few 0.1\,\msun\ in the choice of
$M_{\rm piston}$ value have no effect on the results for the ejecta kinematics (provided the energy deposited
is adjusted to account for the modulation in the binding energy of the material outside of $M_{\rm piston}$).

\begin{figure*}
\epsfig{file=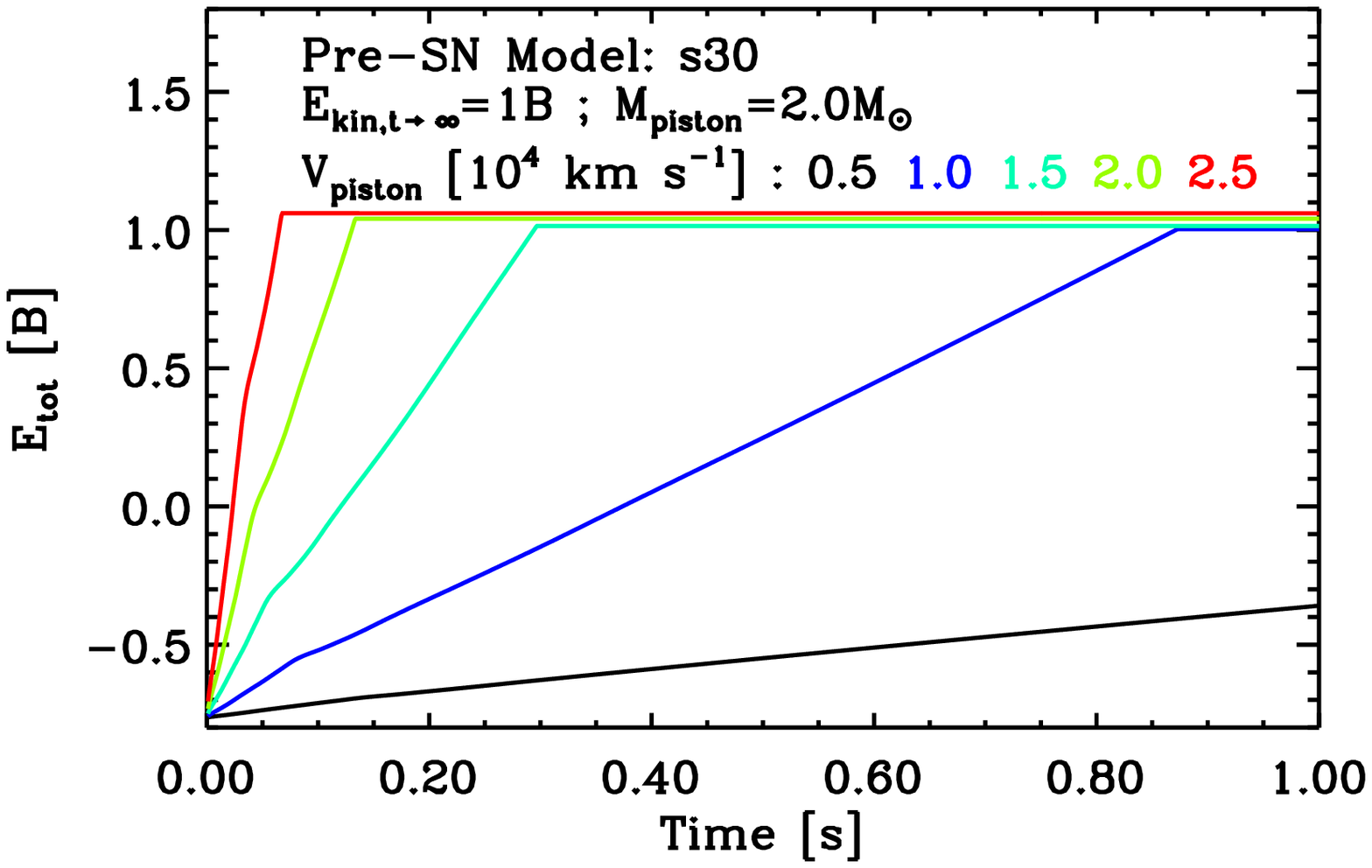,width=8.5cm}
\epsfig{file=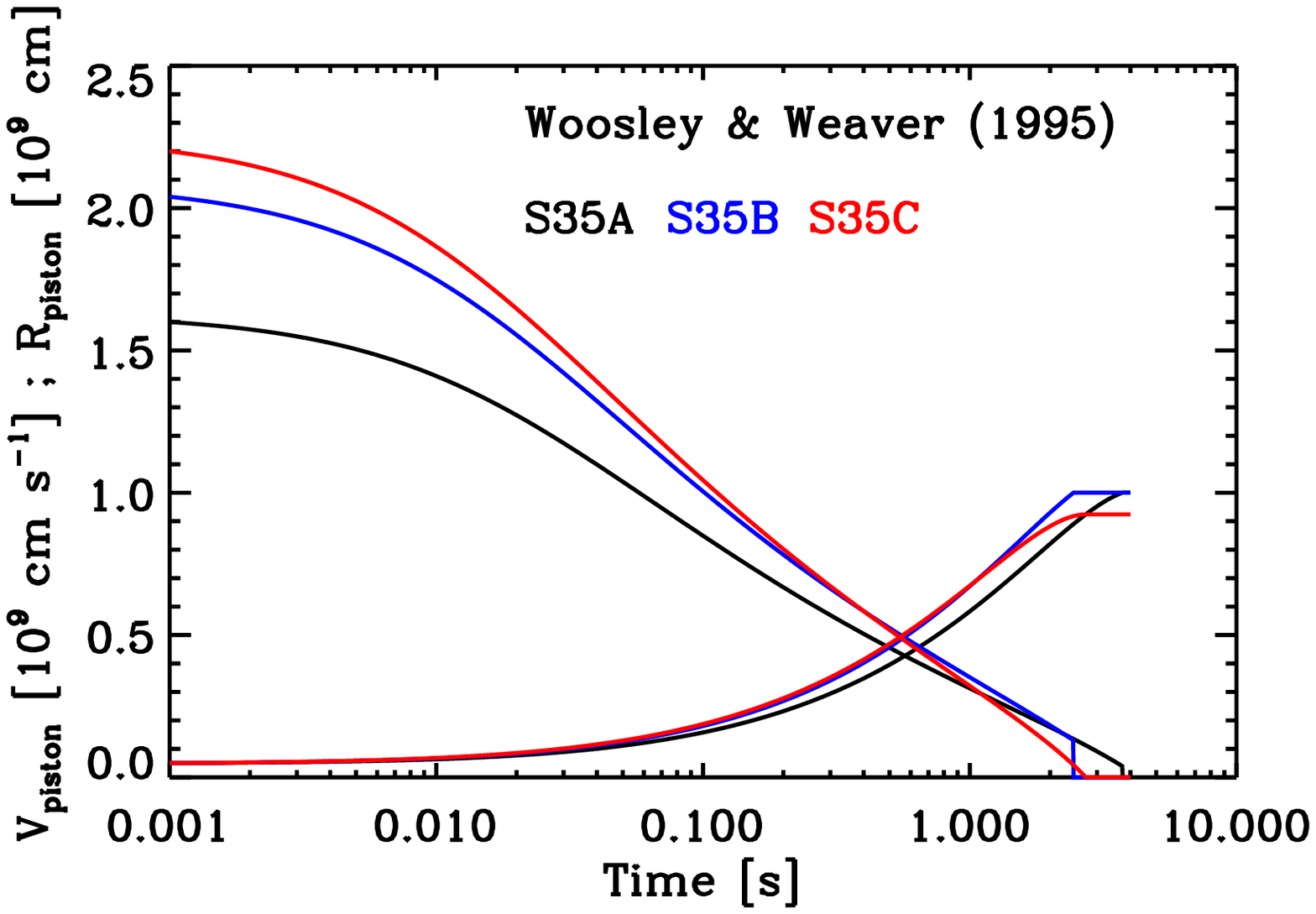,width=8.5cm}
\caption{{\it Left:} Illustration of the evolution of the total ejecta energy in our {\sc v1d} simulations
based on the pre-SN model s30 characterised by $E_{\rm kin}$=1\,B and piston speeds of 5000 (black), 10000 (blue),
15000 (turquoise), 20000 (green), and 25000\,\kms\ (red). For the former model, it requires about 5\,s
to reach the desired ejecta kinetic energy, which is much too long, while for the latter, it takes no more
than 60\,ms, which is much too short. A guess compatible with current state-of-the-art simulations
of core-collapse SN explosions would be on the order of 10000\,\kms.
{\it Right:} Illustration of the piston velocity (downward-pointing curve)
and radius (upward-pointing curve) for the simulations S35A, S35B, and S35C
of \citet{woosley_weaver_95}. [See text for discussion]. \label{fig_piston_ww95}
}
\end{figure*}

Once initiated, the explosion does not occur promptly - the energy is deposited over a finite time.
In their simulation of a 15\,\msun\ star,
\citet{marek_janka_09} find promising signs for a successful neutrino-driven explosion starting at $\sim$500\,ms,
but a few hundred milliseconds or even a full second may be needed to reach an energy
of 1\,B (there is structure in their curves of, e.g. Fig.~9, which makes it difficult to extrapolate in confidence).
In the context of magneto-rotational core-collapse SN explosions, \citet{burrows_etal_07b}
and \citet{dessart_etal_08b} find their
simulations reach $\sim$1\,B about 300\,ms after the shock is launched, while a full second will
be needed to deliver the additional energy to unbind the envelope and boost it
to hypernova energies they seem capable of reaching. In practice, and given all these uncertainties,
we simply adopt a constant piston speed $V_{\rm piston}$, with values of 10000 and 20000\,\kms.
As illustrated in the left panel of Fig.~\ref{fig_piston_ww95}, this choice covers the extremes
from a sudden (energy deposition over $\sim$100\,ms) to a slow-developing explosion (energy deposition over $\sim$1\,s).
Once the piston has delivered the energy aimed for, it is stopped.

The approach of \citet{woosley_weaver_95} is similar but somewhat more complicated. They first position the piston
at some mass cut $M_{\rm cut}$, let it dynamically collapse until a radius of $R_{\rm min}=500$\,km
where it is suddenly set in outward motion with an initial velocity $V_0$. At subsequent times,
the piston trajectory $R(t)$ is given by
$dR/dt = (\alpha G M_{\rm cut}(1/R - 1/R_{\rm min})  + V_0^2 )^\frac{1}{2}$,
until it reaches 10000\,km where it is stopped ($\alpha$ is some adjustable parameter).
With such a trajectory, the bulk of the energy is deposited at early times, when the
piston speed is maximum.
We show in the right panel of Fig.~\ref{fig_piston_ww95} the resulting piston radii and velocities for their simulations
S35A ($V_0=16000$\,\kms, $\alpha=0.50$, and $E_{\rm K\infty}=1.23$\,B),
S35B ($V_0=20400$\,\kms, $\alpha=0.81$, and $E_{\rm K\infty}=1.88$\,B), and
S35C ($V_0=22000$\,\kms, $\alpha=0.95$, and $E_{\rm K\infty}=2.22$\,B), where $E_{\rm K\infty}$
is the asymptotic ejecta kinetic energy.
In these three cases, they obtain remnant masses of 7.38, 3.86, and 2.03\,\msun, respectively.
The last value corresponds to their adopted choice of $M_{\rm cut}$, which thus suggests no fallback
for simulation S35C. The properties of the compact remnant are visibly affected by the explosion energy and the
choice of piston speed. In their study, \citet{woosley_weaver_95}  adopt reference piston speeds $V_0$
in the range 12000 up to 37000\,\kms.
In this work, we find that our results for the remnant mass (see appendix and Tables~2--5)
are comparable to those of \citet{woosley_weaver_95}
for a piston speed of 20000\,\kms, which corresponds to a very short energy-deposition timescale of $\sim$100\,ms
(left panel of Fig.~\ref{fig_piston_ww95}).
Since it seems that the shock revival/powering takes place on a longer timescale, their choice of piston
speeds may be somewhat overestimated (and consequently their remnant masses underestimated).

   As the mass of the core increases from low-mass to high-mass
massive stars, a growing fraction of the helium core eventually falls back. This fallback material remains
dense and hot and limits the progress of the simulation to late times. Hence, all our simulations are halted
at  20000\,s after the piston trigger. Any material in the inner ejecta moving with a velocity smaller
that the local escape speed is trimmed and the simulation is restarted. This rough treatment
likely affects the accuracy of the remnant mass, although our results agree
closely with those of  \citet{woosley_weaver_95} when similar parameters are used. In contrast, for a given
energy deposition, the adopted piston trajectory does not alter sizeably the key ejecta kinematics,
which are the focus of the present study.

\begin{figure*}
\epsfig{file=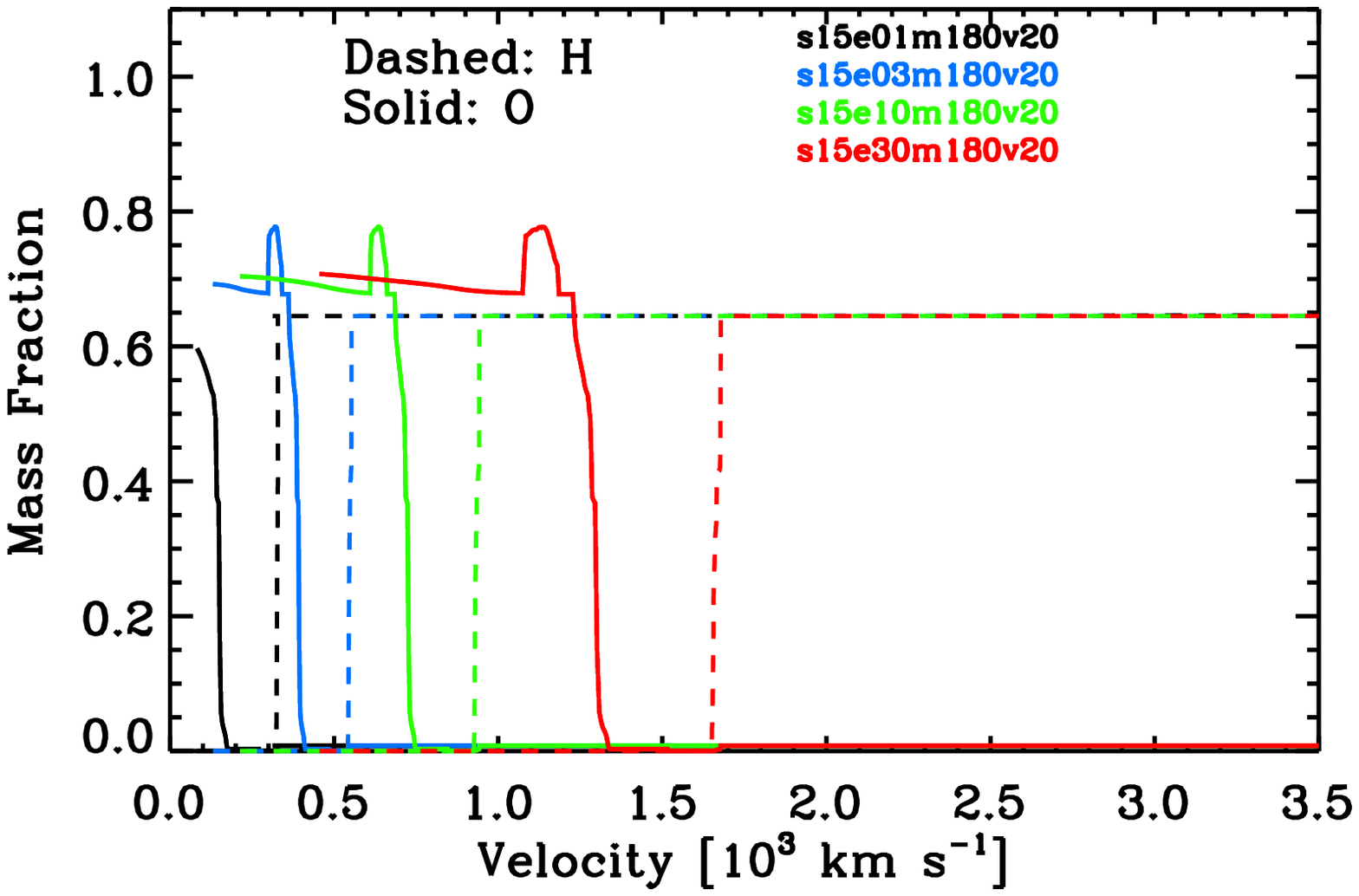,width=8.5cm}
\epsfig{file=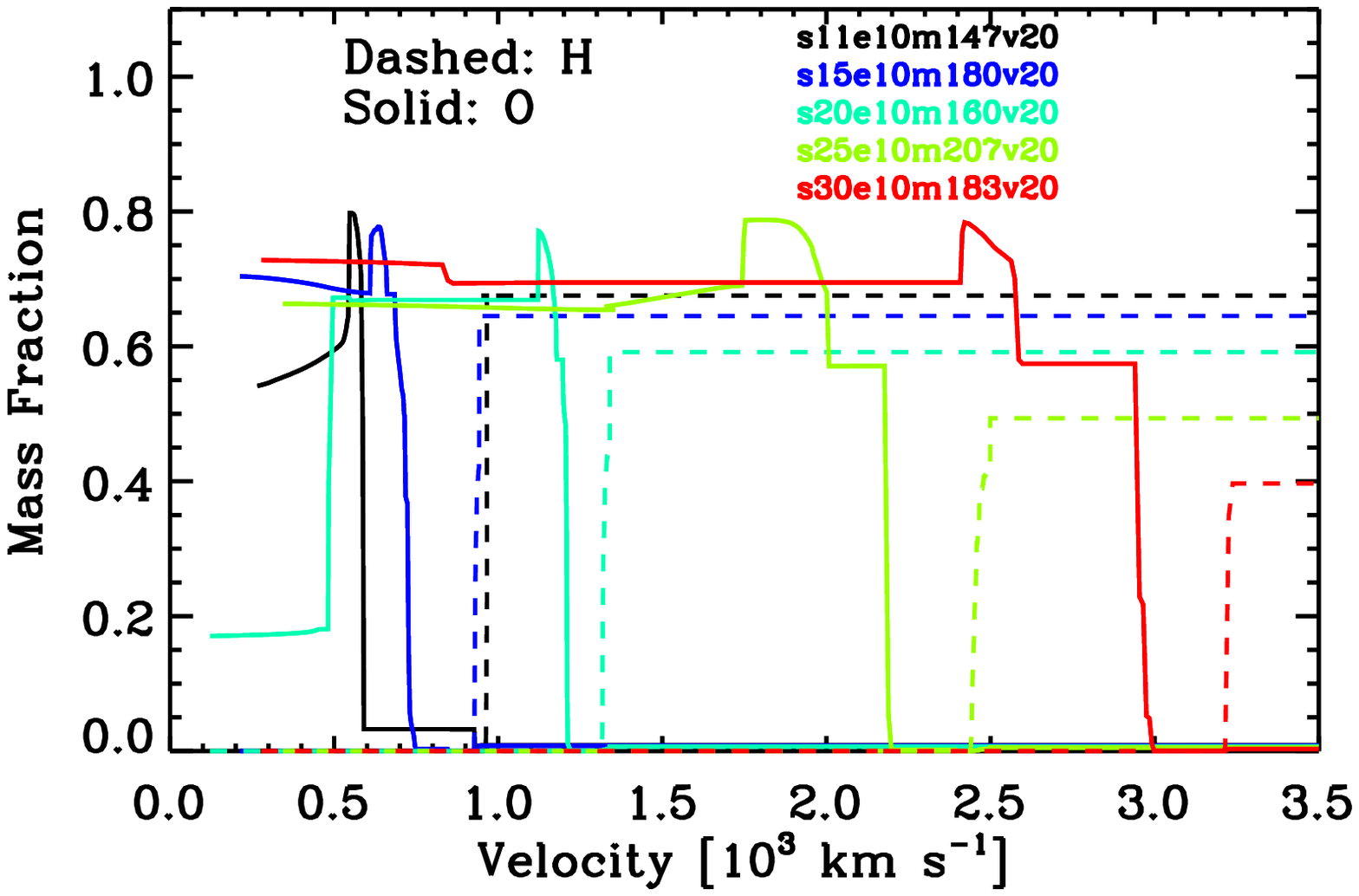,width=8.5cm}
\caption{
{\it Left:} Distribution of hydrogen (dashed line) and oxygen (solid line) with respect to
ejecta velocity for simulations s15e01m180v20 (black), s15e03m180v20 (blue),
s15e10m180v20 (green), and s15e30m180v20 (red), all based on the {\it same pre-SN model} s15.
The time is 70 days after explosion, thus when homology is well established.
Notice the systematic location of oxygen at larger velocities for increasing asymptotic ejecta kinetic energy.
{\it Right:} Same as left, but now showing the results for simulations
s11e10m147v20 (black), s15e10m180v20 (blue), s20e10m160v20 (turquoise),  s25e10m207v20 (green),
and s30e10m183v20 (red), having {\it the same ejecta kinetic energy} of 1\,B but differing in pre-SN model.
The range of velocities we obtain for the outer edge of the oxygen-rich shell is now much larger.
[See text for discussion] \label{fig_comp_vel}}
\end{figure*}

We perform no mixing/smoothing on the progenitor structure, neither prior to nor after the explosion
is launched, and thus do not test the potential (and expected) impact
asymmetric explosions or Rayleigh-Taylor instabilities
would have on observables.

Since the bulk of our discussion is on the ejecta properties, rather than on the emergent radiation,
we do not treat energy deposition from unstable isotopes like $^{56}$Ni. When included,
the associated decay energy is known to lengthen the high-brightness part of the light curve.
This will be important for the computation of detailed non-LTE
time-dependent light curves and spectra using the approach of \citet{DH10},
which is left to a forthcoming paper. Hence, all plateau durations given in Tables~2--6 represent a lower limit.

In the next section, we present results from our simulations. The nomenclature for our models
is such that our simulation called s11e10m147v20 corresponds to one based on the WHW02 pre-SN
model s11, exploded to yield a 1\,B ejecta kinetic energy at infinity (e10), adopting a
piston mass cut at 1.47\,\msun\ (m147) and a piston speed of 20000\,\kms\ (v20).
In practice, we have run simulations for WHW02 pre-SN models s11 up to s30 at every 1\,\msun\ increment
in main-sequence mass (i.e. s11, s12, s13 etc.), but run only simulations for the E10, E12, E15, and E20 rotating
pre-SN models of HLW00. Note also that for low-energy explosions, the fallback material prevents an easy
guess of the asymptotic ejecta kinetic energy (only a fraction of the mass placed on the grid will be kept after trimming
the fallback material, hence changing the total energy on the grid with time).
For ``e01" models (energy deposition to yield a 0.1\,B
ejecta kinetic energy), the simulations ended up asymptotically with an energy as high as $\sim$0.2\,B,
while for more energetic explosions, the two energies agreed within a few percent.
This and numerous other results from our simulations are given in Tables~2--6.

\section{Results}
\label{sect_res}

   In this section, we present the results from the radiation-hydrodynamics simulations with {\sc v1d} and
   based on the pre-SN progenitor models of WHW02 (prefix ``s"; no rotation) and HLW00 (prefix ``E"; with rotation)
   - see Table~\ref{tab_sum_progenitor} for details. A full log of our simulation results is presented in Tables~2--6,
   with additional information given in the appendix. Our exploration is over pre-SN models that differ in their
   main-sequence mass (we adopt the mass range 11--30\,\msun\ for non-rotating models and
   10--20\,\msun\ for rotating models) and piston trajectories that yield asymptotic ejecta kinetic energies
   of 0.1, 0.3, 1.0, and 3.0\,B.

   Our simulations based on such pre-SN models yield successful explosions that would be characterised
   as a II-P event (the progenitor stars are RSGs with a massive hydrogen-rich envelope).
   The general evolution after the piston trigger of the resulting shock-heated envelopes is the following.
   After a delay corresponding to the shock-crossing
   time through the envelope (which we denote $t_{\rm SBO}$), the ejecta expand
   and accelerate for a few days until reaching a coasting phase in which all mass parcels
   move at a constant velocity.
   Expansion causes dilution and a decrease of the ejecta optical depth
   as the inverse square of the time, reduced further by the recombination of ions to their neutral state.
   While the photosphere (defined as the location where the inward-integrated continuum optical depth equals 2/3)
   moves initially outward in radius, within about 30--50\,d of shock-breakout it stabilises
   at a radius of $\sim$10$^{15}$\,cm and gives a plateau to the SN light curve \citep{dessart_etal_08}.
   This ``photospheric" phase
   ends when the ejecta become optically-thin in the continuum. While this scenario holds qualitatively in all the
   simulations we performed for this work, differences in the pre-SN models and/or the adopted
   explosion parameters can lead to significant quantitative differences. A number of studies have focused on the
   resulting properties of the emergent light, such as recently \citet{kasen_woosley_09}, and are thus not presented
   in detail here, although we give a comprehensive summary of our results in the Appendix and Tables~2--6.
   However, since this has not been discussed in detail elsewhere,
   we focus here on the properties of the SN ejecta kinematics and chemical stratification.
   Ultimately, our goal is to identify spectroscopic observables that can complement the restricted information
   provided by the light curve in order to better constrain the properties of the progenitor star and the explosion physics.

\begin{figure}
\epsfig{file=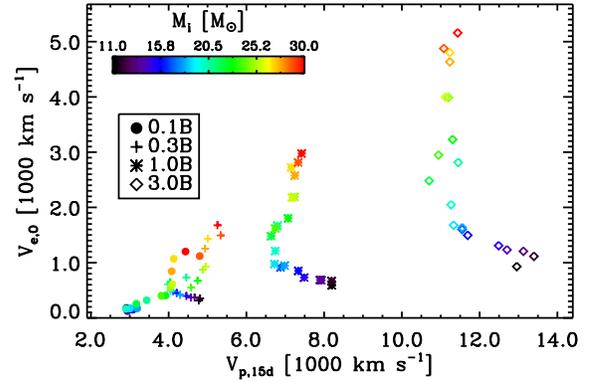,width=8.5cm}
\caption{Distribution of the velocity at the outer edge of the oxygen-rich shell
$V_{\rm e,O}$ as a function of the ejecta photospheric velocity at 15\,d after
shock-breakout $V_{\rm p,15d}$ for our {\sc v1d} simulations adopting a piston speed
of 20000\,\kms, and set up to yield
an asymptotic ejecta kinetic energy of 0.1\,B (dots), 0.3\,B (pluses), 1\,B (asterisks), and 3\,B (diamonds).
We employ a colour-coding to give the main-sequence mass for each model.
With this nomenclature, the model s11e01m147v20 appears as a black dot, and
the model s30e30m183v20 as a red diamond.
Comprehensive results from our simulations are given in Tables~2--5. \label{fig_summary_vel_norot}}
\end{figure}

   \subsection{Results based on non-rotating pre-SN models}

   For a given progenitor mass, increasing the energy deposited by the piston yields an ejecta with a higher kinetic energy
   at infinity and a larger velocity in each mass shell. With ejecta kinetic energy of $\sim$0.1, 0.3, 1.0, and
   3.0\,B, simulations s15e01m80v20, s15e03m80v20, s15e10m80v20, and s15e30m80v20 show
   a maximum velocity of the outer-edge of the oxygen shell $V_{\rm e,O}$  at 155, 396, 730, 1308\,\kms, hence encompassing
   about a factor of ten in this set corresponding to the pre-SN model s15 (left panel of Fig.~\ref{fig_comp_vel}).
   The velocity at the inner edge of the hydrogen-rich envelope $V_{\rm i,H}$, which is located just outside of the oxygen-rich shell
   in the pre-SN progenitor envelope (the two are separated by a narrow helium-rich shell),
   is generally a few 100\,\kms\ larger and thus does not require a specific discussion.
   But if we fix the energy deposition to yield an ejecta kinetic energy of 1\,B and now 
   vary the main-sequence mass of the pre-SN model,
    from 11 to 30\,\msun, the range of values for $V_{\rm e,O}$ is considerably enhanced.
    In the right panel of Fig.~\ref{fig_comp_vel}, we show such results for our simulations
    s11e10m147v20, s15e10m180v20, s20e10m160v20, s25e10m207v20, and s30e10m183v20,
    which give values of $V_{\rm e,O}$ now stretching from 590 to 2975\,\kms.
    The larger size of the helium-core for larger main-sequence mass stars produce SN ejecta that
    eject core-embedded oxygen-rich material at a larger velocity for a {\it given} ejecta kinetic energy.
    This stems naturally from the fact
    that in larger mass progenitors, the bigger helium core represents a larger fraction of the total mass, i.e.
    the outer oxygen-rich shell is located closer to the progenitor surface.
    We stress here that this result depends primarily on the {\it main-sequence} mass, since it
    determines the mass of the helium core.
    This is somewhat paradoxical because the SN radiation
    is generally interpreted in terms of the properties of the star at the time of collapse. In SNe II,
    mass loss does not alter the mass of the helium core, and the connection to the main-sequence
    mass of the progenitor can therefore be made. The final mass at collapse does, however, enter the problem as it sets
    the magnitude of the mass-weighted mean ejecta velocity. However, our set of pre-SN models shows a very narrow range of
    values for the final mass so that the trend observed here is controlled primarily by the large variation in $M_{\rm e,O}/M_{\rm f}$
    rather than the modest variation in $M_{\rm f}$.\footnote{The set of pre-SN stellar-evolution models from WHW02
    and HLW00 were computed with specific prescriptions for
    the treatment of mass loss etc. The trends we identify hence apply to these pre-SN models and would
    be altered if, for example, we were to take Population {\sc iii} star models.}

     These results suggest that in SN II-P ejecta, both explosion energy and progenitor main-sequence mass correlate with
     the  asymptotic value of $V_{\rm e,O}$. The explosion energy can be determined independently, as it
     affects the mean velocity of the ejecta, which can be inferred through an estimate of the photospheric velocity
     at some given time.
     Here, we use the ejecta photospheric velocity at 15\,d after shock breakout, $V_{\rm p,15d}$. This measurement
     can be done using spectroscopic observations and fitting hydrogen Balmer lines
     by means of radiative-transfer simulations \citep{DH05_qs_SN}.
     Alternatively, one can use the photospheric velocity at 50\,d after explosion ($V_{\rm p,50d}$), inferred through a measurement
     of the Doppler-velocity corresponding to maximum P-Cygni-profile absorption in Fe{\sc ii}\,5169\AA\
     \citep{DH05_epm,kasen_woosley_09,DH10}.
     In Fig.~\ref{fig_summary_vel_norot}, we now show the resulting distribution of values of $V_{\rm e,O}$
     versus $V_{\rm p,15d}$ for a large set of simulations based on the non-rotating pre-SN models of WHW02.
     A colour-coding is used to differentiate the initial (main-sequence) mass of each model, while symbols
     are used to differentiate the asymptotic kinetic energy aimed for (dots: 0.1\,B; pluses: 0.3\,B; asterisks: 1.0\,B;
     diamonds: 3.0\,B). Models of distinct ejecta-kinetic energy are now well separated, provided this energy
     is $\sim$0.3\,B or larger. High-energy explosions of 3.0\,B are characterised by $V_{\rm p,15d}\gtrsim$10000\,\kms,
     with values of $V_{\rm e,O}$ ranging from $\sim$1000\,\kms\ up to 5000\,\kms. For the more standard
     explosion energy of 1.0\,B, $V_{\rm p,15d}\sim$7000\,\kms, and values of $V_{\rm e,O}$ range from
     500\,\kms\ up to a lower maximum of 3000\,\kms. For yet lower explosion energies, values of $V_{\rm e,O}$
     and  $V_{\rm p,15d}$ continue to decrease and start to overlap.\footnote{As the energy is lowered,
     the distributions become more compact and a unique association of $V_{\rm e,O}$ and $V_{\rm p,15d}$ is difficult.
     This results in part from a problem in our simulations of low-energy explosions for higher mass stars,  which
     are all characterised by a  large fallback mass. In these simulations, the fallback mass takes away binding energy and leads
     to overestimated energies for the ejected material. In such under-energetic explosions, hardly any core-embedded oxygen-rich material is ejected,
     making the discussion of $V_{\rm e,O}$ somewhat irrelevant in this situation.} However, for explosion energies of 0.3\,B or less,
     values of $V_{\rm e,O}$ are systematically below 2000\,\kms, while values of $V_{\rm p,15d}$ never
     exceed 6000\,\kms.
     Note that the kink in the distribution of data-points in Fig.~\ref{fig_summary_vel_norot} for initial masses
     of $\sim$20\,\msun\ corresponds to the pre-SN models in our sample that have the largest mass at the time of collapse,
     hence the lowest mean-ejecta velocity for a given explosion energy (they have the lowest kinetic energy per unit mass).

\begin{figure}
\epsfig{file=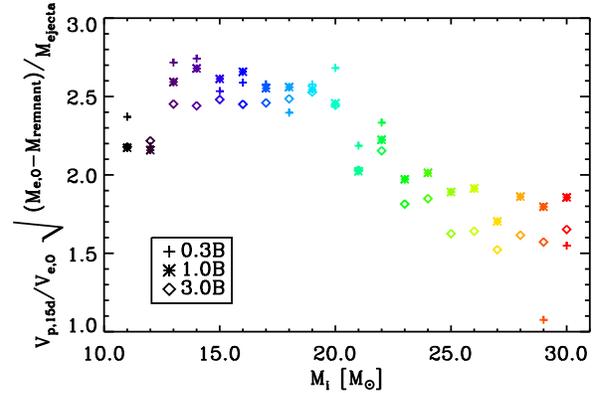,width=8.5cm}
\caption{Illustration of the variation of the quantity
$\frac{V_{\rm p,15d}}{V_{\rm e,O}} \sqrt{\frac{M_{\rm e,O}-M_{\rm remnant}}{M_{\rm ejecta}}}$
versus initial mass $M_{\rm i}$ for our {\sc v1d} simulations based on the non-rotating pre-SN models  of WHW02
and exploded to yield an asymptotic ejecta kinetic energy of 0.3 (pluses), 1.0 (asterisks), or 3.0\,B (diamonds;
the symbol and colour assignment is the same as that used in Fig.~\ref{fig_summary_vel_norot}). Note that we exclude from this figure
the simulations that use an explosion energy of 0.1\,B since in this situation, the fallback magnitude is huge and
no core-embedded oxygen-rich material is ejected ($M_{\rm e,O}<M_{\rm remnant}$).
This issue also applies to a few simulations with
an explosion energy of 0.3\,B for pre-SN models with a main-sequence mass between 23 and 28\,\msun.
[See text for discussion]. \label{fig_vcheck}}
\end{figure}

     The apparent scatter of data-points shown in Fig.~\ref{fig_summary_vel_norot}, both with explosion energy and pre-SN model,
     can be reduced by inspecting dependent variables.
     In our set of simulations, we recover the scaling of the ejecta velocity with the square-root of energy over mass.
     For example, we find that the ratio of $V_{\rm e,O}$ and $\sqrt{e_{\rm kt}/(M_{\rm e,O}-M_{\rm remnant})}$ is independent
     of the ejecta kinetic energy $e_{\rm kt}$ (note, however, that this ratio varies significantly with pre-SN model, reflecting variations in $M_{\rm e,O}/M_{\rm f}$).
     Here, $V_{\rm e,O}$ is a representative ejection velocity for the
     core-embedded oxygen-rich material, whose mass scales with  $M_{\rm e,O}-M_{\rm remnant}$.
     The above scaling holds provided the fallback is never too large to yield $M_{\rm e,O} < M_{\rm remnant}$,
     a circumstance that characterises under-energetic explosions.
     We also find that $V_{\rm p,15d}$ over $\sqrt{e_{\rm kt}/M_{\rm ejecta}}$ is independent of explosion energy.
     The quantity  $X=\frac{V_{\rm p,15d}}{V_{\rm e,O}} \sqrt{\frac{M_{\rm e,O}-M_{\rm remnant}}{M_{\rm ejecta}}}$,
     is thus independent of energy. As shown in Fig.~\ref{fig_vcheck} (our {\sc v1d} simulations with explosion energy of 0.1\,B
     experience complete fallback of the core-embedded oxygen-rich material and are thus not plotted in this figure),
     simulations for a given pre-SN model but different explosion energies show $X$ values
     that are generally within $\sim$10\% of each other.

     The uniformity of values of $V_{\rm p,15d}$ for a given explosion energy but a range of pre-SN models
     (Fig.~\ref{fig_summary_vel_norot}) suggests it is a rough but easy
     diagnostic of the explosion energy characterising a SN II-P. The large range in progenitor main-sequence mass for
     our set of pre-SN models
     also suggests that it should not be dramatically altered by uncertainties in stellar evolution and thus represents
     a valuable guide (unless the mass-loss rate prescriptions used in the pre-SN stellar evolutionary calculations are way off).
     Having estimated the explosion energy through a measurement of $V_{\rm p,15d}$, the value of
     $V_{\rm e,O}$ can place tight constraints on the progenitor main-sequence mass. Our results suggest that
     main-sequence stars less massive
     that $\sim$20\,\msun\ tend to produce SN II-P ejecta with values of $V_{\rm e,O}$ smaller than 500, 1000, and
     2000\,\kms, for explosion energies of 0.3, 1.0, and 3.0\,B, respectively.
     For any explosion energy in the range 0.1 to 3.0\,B, simulations based on the s11 pre-SN model show values of  $V_{\rm e,O}$
     that are systematically below 1000\,\kms\ while those based on the s30 pre-SN model show values
     always in excess of 1000\,\kms\ no matter what the explosion energy is.
     The maximum speed of the oxygen-rich shell sets therefore a very strong constraint on the progenitor mass of
     SNe II-P, distinguishing between a low-mass or a high-mass massive star as the progenitor.

     The above simulations employ a piston speed of 20000\,\kms. As shown in Tables~2--6,
     employing a piston speed of 10000\,\kms\ for the same set of pre-SN models does not in general alter by more than $\sim$10\%
     the values previously found (one exception is when the lowering of the piston speed leads to fallback of the entire
     helium core, as happens for model E20e10m200v10 compared to model E20e10m200v20).
     This suggests that, in the present (1D) context, the ejecta kinematics are very sensitive to the explosion energy,
     but not to the details of the explosion mechanism nor the way we numerically explode these pre-SN stars
     (e.g. the exact value of $M_{\rm piston}$).\footnote{As
     we discuss in the appendix, this does not apply to the resulting remnant mass, which we find to be dramatically
     sensitive to the piston speed.}

\begin{figure}
\epsfig{file=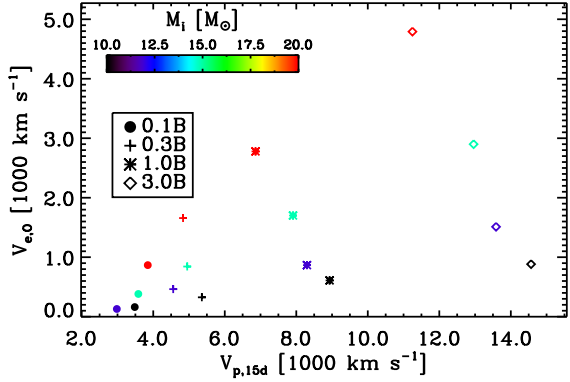,width=8.5cm}
\caption{Same as Fig.~\ref{fig_summary_vel_norot}, but now for our simulations based on a set of rotating pre-SN models.
Comprehensive results from our simulations are given in Table~6. [See text for discussion]. \label{fig_summary_vel_rot}}
\end{figure}

   \subsection{Results based on rotating pre-SN models}

    As emphasised above, the helium-core mass is a growing function of main-sequence mass. However, stellar
    rotation breaks this simple correlation, as it tends to increase the helium-core mass for a {\it given} main-sequence
    mass. We then anticipate that the values of $V_{\rm e,O}$ presented above for non-rotating progenitors
    place an {\it upper-limit} on the main-sequence mass of a given SN progenitor.
    We thus performed another set of simulations, using the same procedure as above for the non-rotating models,
    but now employing the rotating pre-SN models E10, E12, E15, and E20 of HLW00 (a census of results is given
    in Table~6). As shown in Fig.~\ref{fig_summary_vel_rot}, the trend identified for non-rotating progenitors
    holds qualitatively, but the results are quantitatively different. Going from 0.1, 0.3, 1.0, to 3.0\,B explosion energies, we
    obtain increasing mean values of  $V_{\rm p,15d}$ of about 3000, 5000, 8000, and 14000\,\kms, much larger
    than previously obtained. As expected,
    for a given main-sequence mass and explosion energy, the oxygen ejected in the SN II-P simulation
    reaches maximum speeds that are larger for rotating pre-SN models.
    For example, comparing the results for pre-SN
    models s15 and E15, we obtain $V_{\rm e,O}$ of 730 and 1701\,\kms, respectively (simulations s15e10m180v20
    and E15e10m176v20). While a main-sequence mass limit of $\sim$20\,\msun\  emerged for the representative value
    $V_{\rm e,O}\sim$1000\,\kms\ in 1.0\,B explosions of non-rotating SNe II-P progenitors, the allowance for rotation lowers
    this mass limit to $\sim$12\,\msun.
    Note that we are here addressing how rotation conditions the helium-core mass of a
    massive star and the signatures it leaves in a SN II-P ejecta; we are not considering how it
    may play a role in the explosion mechanism itself or how that initial angular momentum is eventually
    distributed in the star at the time of collapse.

\section{Discussion and conclusions}
\label{sect_concl}

\begin{figure*}
\epsfig{file=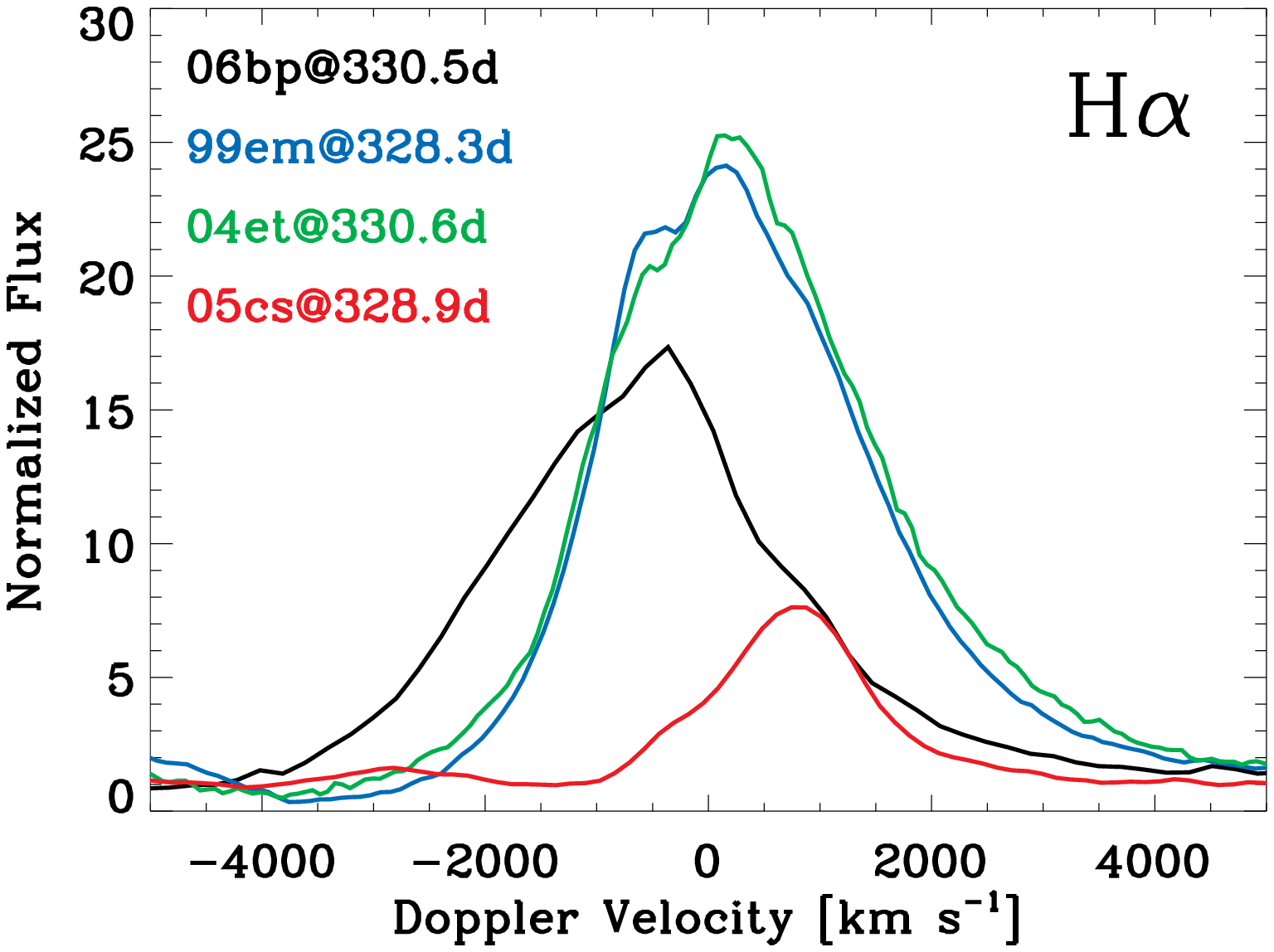,width=8.55cm}
\epsfig{file=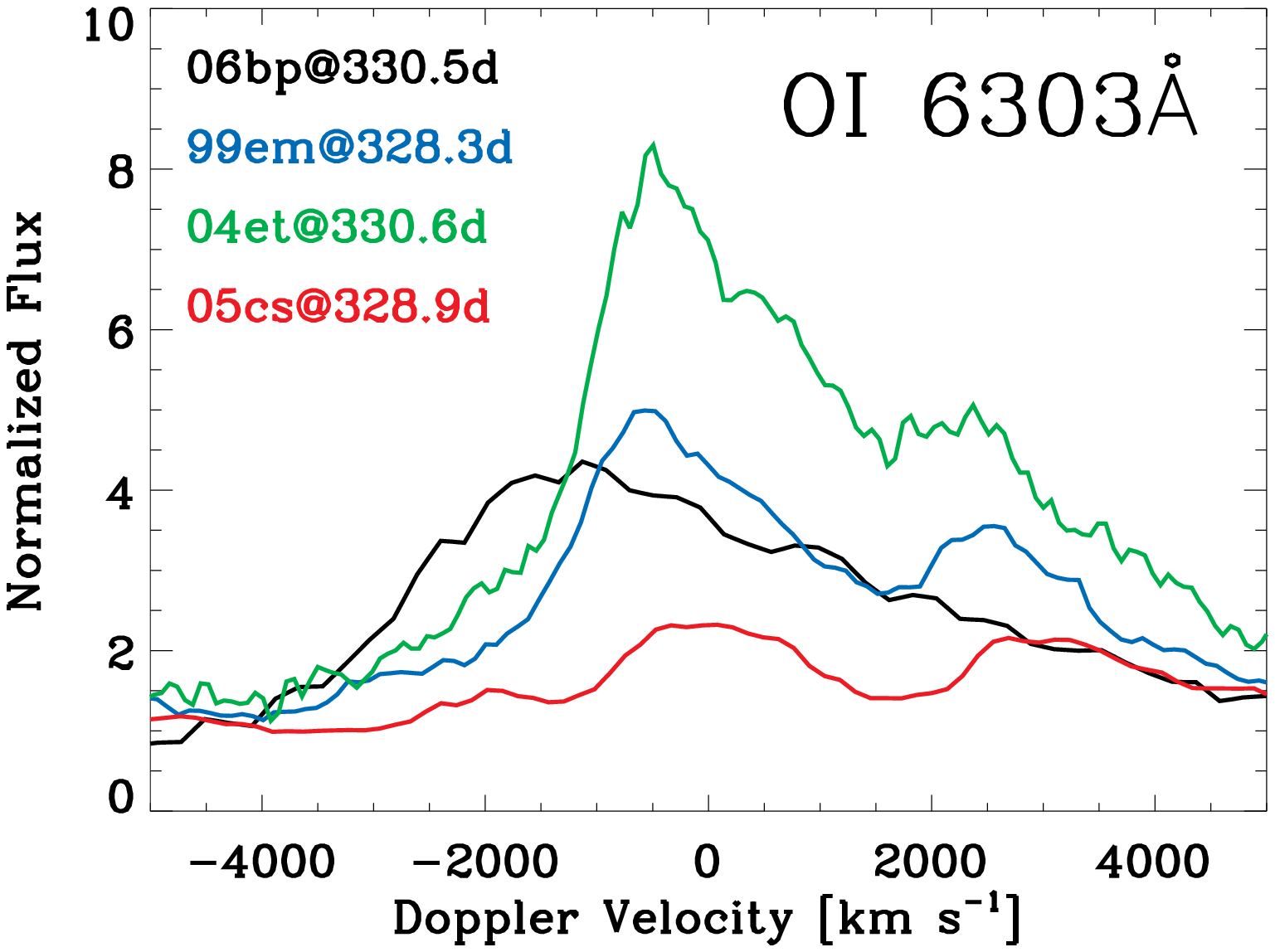,width=8.5cm}
\caption{Illustration of the H$\alpha$ (left) and O{\sc i}\,6303--6363\AA\ (right) line profiles versus Doppler velocity
for four SNe II-P at about 300--350\,d after explosion.
Specifically, we show data for SN1999em (blue; \citealt{leonard_etal_02a}; age of $\sim$328 days)
SN2004et (green; \citealt{sahu_etal_06}; age of $\sim$303\,d), SN2005cs (red; \citealt{pastorello_etal_09}; age of $\sim$329 days)
and SN2006bp (black; \citealt{quimby_etal_07}; age of $\sim$330\,d). The data has been de-redshifted using
the redshift values quoted in those works, and shifted vertically to best render the profile widths.
[See text for discussion]. \label{fig_vel_obs}
}
\end{figure*}

In this paper, we have presented results from radiation-hydrodynamics simulations of core-collapse SN explosions,
using two sets of pre-SN progenitor models {\it evolved at solar metallicity} and accounting or not for stellar rotation
(WHW02, HLW00). Our sample of objects includes stars
that possess a sizeable hydrogen-rich envelope, die in their RSG stage, and produce a SN II-P.
Because of mass loss, their final mass at the time of collapse is expected to be fairly degenerate
although somewhat uncertain ($\lesssim$15\,\msun; WHW02).
However, an important property of stellar structure and evolution is that stars with larger main-sequence mass
possess more massive helium cores. The range in the non-rotating pre-SN models of WHW02 we employ is
1.75 to 9.18\,\msun\ for main-sequence masses 11 to 30\,\msun. Rotating stars achieve the same helium-core
mass for lower main-sequence mass, so that a given helium-core mass can be used to set an upper limit to the main-sequence
star mass. Because the objects in our sample have not been peeled off down to the helium core, the trend of increasing
helium-core mass with increasing main-sequence mass is preserved. In this study, we quantified the asymptotic
kinematic properties of  SN II-P ejecta, and in particular searched for trends that would distinguish objects
with a markedly different helium-core mass.

With our radiation-hydrodynamics simulations, we recover the trivial result that
more energetic explosions lead to larger ejecta expansion rates, as visible from the velocities of various
shells of the progenitor envelope, or by inspection of the photospheric velocity.
However, we emphasise that for a {\it given} ejecta kinetic energy,
pre-SN stars with a higher {\it main-sequence} mass yield SN ejecta with similar photospheric velocities
but increasing velocities of helium-core material. One important element abundant in the helium-core is oxygen,
and one important location is the outer edge of the oxygen-rich shell. For a standard-energy core-collapse SN
explosion of 1\,B, we find that the velocity of the outer edge of the oxygen-rich shell is smaller than $\sim$1000\,\kms\
for  pre-SN stars with main-sequence mass smaller than $\sim$20\,\msun. Our {\sc v1d} simulations based on {\it rotating}
pre-SN models suggest that for the same ejecta kinetic energy,  the same oxygen ejection speeds can be reached
but for pre-SN stars with a lower main-sequence mass (for the last example, the mass threshold may be reduced down
to $\sim$15\,\msun). All these results are shown in Figs.~\ref{fig_summary_vel_norot} and \ref{fig_summary_vel_rot} and
given in Tables~2--6.
We find that tight constraints can be placed on the main-sequence mass of
SN II-P progenitors based on two measurements, one of the photospheric velocity, which testifies for the
strength of the explosion, and one of the velocity of the outer edge of the oxygen-rich shell, which
testifies for the mass of the helium core and thus that of the star on the main sequence.
Although potentially difficult to assess accurately on a case-by-case basis, the correlations identified above should emerge
in observations of a {\it statistical} sample of SNe II-P in which under-energetic events are excluded to ensure some homogeneity.
Combined with inferences based on light-curve modelling, which is sensitive to the properties of the hydrogen-envelope,
but not to those of the helium core, one may better constrain the properties of the progenitor and of the explosion.

Inferences on ejecta kinematics can be done from line-profile morphology in spectroscopic observations.
During the photospheric-phase of the SN, such measurements may be complicated by effects associated
with line overlap, optical-depth \citep{DH05_epm}, time-dependent effects \citep{DH08_time}, ionisation (different lines are
seen at distinct epochs, and these lines may be differentially affected by the previous two effects),
or more generally the peculiarity of line-profile formation in SNe II \citep{DH05_qs_SN}.
For example, Balmer lines in the early-time spectra of SN1987A yield a maximum P-Cygni absorption corresponding to
a Doppler velocity that
overestimates by up to a factor of two the contemporaneous photospheric velocity \citep{DH10}.
The ejecta kinetic energy of a SN II-P is best estimated at such early times when the photosphere
is within the shock-heated hydrogen-rich part of the ejecta. The choice of 15\,d after explosion, as above,
is good -  a later time during the photospheric phase is an alternative but the measurement may then
be influenced by multi-dimensional effects and decay heating.
At late times in the nebular phase of the SN, such uncertainties are gone and one can simply
measure the width of a line to assess the expansion velocity of the corresponding emitting region
(this is best done for singlets and in the absence of line overlap). Such measurements have
been performed on O{\sc i}\,6303--6363\AA\ in SNe Ib/c to address the explosion energy and morphology,
as well as the oxygen mass ejected \citep{modjaz_etal_08,tanaka_etal_09,taubenberger_etal_09,
milisavljevic_etal_10,maurer_etal_10}.

Unfortunately, there exists only scant data on late-time spectra of SNe II-P, in particular extending out
to 300 days after explosion, i.e. when the O{\sc i}\,6303--6363\AA\ is well developed.
These objects include SN1999em \citep{leonard_etal_02a},
SN2004et (CfA archive; age of $\sim$303\,d), SN2005cs \citep[age of $\sim$329 days]{pastorello_etal_09}
and SN2006bp \citep[age of$\sim$330\,d]{quimby_etal_07}, for which we show in Fig.~\ref{fig_vel_obs}
the observed (but rest-frame) H$\alpha$ (left) and O{\sc i}\,6303--6363\AA\ (right) lines.
The O{\sc i} line is a doublet whose components appear well separated and narrower than the corresponding
H$\alpha$ line, a result that is expected from the chemical stratification of the progenitor.
Note however that if the explosion morphology was highly asymmetric, this relationship could be broken.
We also note that the widths of the observed H$\alpha$ line varies from small (SN2005cs) to large (SNe 1999em and 2004et)
and very large (SN2006bp). While H$\alpha$ appears quite symmetric with respect to line centre in SNe1999em
and 2004et, it is skewed to the blue in SN2006bp and to the red in SN2005cs.

Photospheric velocities of SNe 2005cs, 1999em, and 2006bp at 15 days after explosion have been inferred to be
about 4700, 8800, and 10300\,\kms\ \citep{DH06_SN1999em,dessart_etal_08}. Our simulations then suggest ejecta kinetic
energies of $\sim$0.3\,B (\citealt{UC_08} obtain an explosion energy of $\sim$0.4\,B),
$\gtrsim$1.0\,B (\citealt{utrobin_07} obtains an explosion energy of $\sim$1.3\,B),
and $\sim$2.0\,B, respectively. \citet{sahu_etal_06} find that the spectral line profiles
of SN2004et are somewhat larger than those of SN1999em at similar dates during their photospheric phase, which suggests
that its explosion energy is larger than 1.0\,B (\citealt{utrobin_chugai_09} obtain an ejecta kinetic energy of $\sim$2.3\,B).
Our estimates of the explosion energy based on $V_{\rm p,15d}$ alone is in good agreement with, although not a replacement of,
such tailored radiation-hydrodynamics simulations.
Using Figs.~\ref{fig_summary_vel_norot} and \ref{fig_summary_vel_rot}, together with the observations of the 
O{\sc i}\,6303-6363\AA\  line width as a guide, we can estimate a representative main-sequence mass of the progenitor.
For SN2005cs, the low explosion energy prevents a good estimate
although the very narrow half-width-at-half-maximum\footnote{
An alternative is to consider the half-width at zero line flux (HWZF), in which case one obtains
a larger estimate for the oxygen ejection speed and thus a larger estimate of the progenitor mass.
However, this measurement is both imprecise and inaccurate.
First, it is difficult to locate where the O\,{\sc i} line flux goes to zero.
Second, in such SNe II-P ejecta at $\sim$300\,d after explosion, the O\,{\sc i}\,6303--6363\AA\ region overlaps
with Fe{\sc ii} (as well as Fe\,{\sc i}) lines so that the flux seen in the wing of the O\,{\sc i}\,6303\AA\ line
is corrupted by Fe{\sc ii}-line flux (Dessart \& Hillier, to be submitted). 
The easily-measured HWHM gets the bulk of the O\,{\sc i}-line flux and represents a diagnostic less prone to errors.  
Furthermore, the HWZF will be relatively more sensitive to the effects of mixing, which leads to a spatial redistribution of the oxygen. 
Stronger mixing would likely produce a larger value of the HWZF, even for pre-SN progenitors with
identical helium-cores} (HWHM) of $\sim$700\,\kms\ suggests the progenitor cannot be more massive than
about 20\,\msun. For SN1999em, the HWHM of $\sim$1000\,\kms\ suggests a progenitor main-sequence mass on the order of 18\,\msun.
For SN2004et, since it has a higher ejecta kinetic energy but a similar O{\sc i} line-profile shape would suggest that its progenitor
main-sequence mass is less than $\sim$18\,\msun.
For SN2006bp, the skewness of the profile (both for O{\sc i} and H$\alpha$) suggests that the explosion may have been
quite asymmetric or that dust is forming, scattering away the photons emitted from the receding part of the ejecta.
From the large width of the O{\sc i} line, we can however exclude a low-mass massive star as the progenitor.
These predictions on main-sequence mass are compatible with those obtained from hydrodynamical simulations for
SNe 2005cs and 1999em, but the relatively narrow O{\sc i}\,6303\AA\ line in the nebular spectrum of SN2004et argues against
the high main-sequence mass proposed by \citet{utrobin_chugai_09}.\footnote{Given the analogous H$\alpha$ and O{\sc i}
profile shapes observed in SNe 1999em and 2004et, and the larger explosion energy inferred for the latter,
our simulations support a lower or a comparable main-sequence mass  for the SN2004et progenitor. Based exclusively on light-curve and photospheric-velocity modelling, \citet{utrobin_chugai_09} propose a larger main-sequence mass of 25--29\,\msun,
a much larger ejecta mass of 24.5\,\msun\ (more than twice as large as our value for our simulations based
on the s25--s30 pre-SN models,  suggesting a surprisingly weak stellar-wind mass loss),
a dense core of $\sim$2.5\,\msun\ (their Fig.~1; note that the mass density should remain large
out to the edge of the helium core at 8.1\,\msun, rather than dropping at a value of $\sim$2.5\,\msun).
From Fig.~\ref{fig_summary_vel_norot} and this set of $E_{\rm kin}$ and $M_{\rm i}$,
one reads a value of $V_{\rm e,O}$ on the order of 2500--3000\,\kms, which
appears too large for the observed O{\sc i} line. Their large initial-mass estimate is incompatible with our set of models for
rotating massive stars evolved at solar metallicity, which explode as SNe Ib/c for $M_{\rm i} \gtrsim 20$\,\msun.}

Obviously, these inferences are not trivial and will require detailed radiative-transfer modelling. At present, given the scarcity
of nebular-phase SNe II-P spectra, it seems difficult, and overly ambitious, to evaluate the accuracy of the method we present.
However, we can identify two sources of uncertainty. The first is inherent to the simulations presented in this work and concerns
the adequacy of the pre-SN models (e.g., with respect to mass loss),
the potential role of multi-dimensional effects caused by an aspherical explosion
and the mixing associated with Rayleigh-Taylor instabilities, the presence of stellar rotation etc.
Inspection of Figs.~\ref{fig_summary_vel_norot} and \ref{fig_summary_vel_rot}
gives some measure of the sensitivity and uncertainty of the main-sequence mass corresponding
to a set of $V_{\rm e,O}$ and $V_{\rm p,15d}$ values. In the quoted mass estimates above, an uncertainty of a few solar masses seems
to apply. Allowance for rotation will lead to a revision downward of the mass inferred from non-rotating models.
The second is the validity of considering the width of the O{\sc i}\,6303--6363\AA\ doublet line as representative of the speed of
the core-embedded oxygen-rich material ejected.\footnote{We note that the minimum velocity of the
hydrogen material inferred from H{\sc i} Balmer lines could be used in combination to this
O{\sc i}-line measurement to make the interpretation more secure.}
This set of forbidden lines forms through collisional excitation and radiative
de-excitation, a process that prevails when the ejecta density becomes on the order of 10$^8$\,cm$^{-3}$.
In their simulation of SN1987A, \citet{KF98a,KF98b} find that this line is thermally excited at nebular times up to $\sim$500\,d,
and that it forms in the inner regions of the ejecta rich in Si/Ne/Mg/O/C. If this applies to SNe II-P at $\sim$300\,d too, it suggests
that the O{\sc i} line may not be strongly sensitive to the distribution and
amount of $^{56}$Ni, and that it should represent
with fidelity the oxygen that was at the time of collapse in the helium core.
We are aware of these complicated and important physical issues, but in this paper, we merely wish to
emphasise the stiff dependence of $V_{\rm e,O}$ with $M_{\rm i}$,
which, intuitively, should produce a clear signature in the width of the nebular phase O{\sc i}\,6303--6363\AA\ line.
This correlation is an important part of the characterisation of SNe II-P progenitor and explosion properties,
which should complement the inferences based on the modelling of the light curve and the photospheric velocity -
it has so far been ignored.
Importantly, more nebular-phase optical spectroscopic data of SNe II-P have to be gathered to build a statistical sample
and investigate kinematics of the oxygen-rich ejected material.
Furthermore, such a spectroscopic dataset would allow us to address the departures from spherical symmetry in the inner
ejecta and the cumulative mass of oxygen ejected. Together with the knowledge of the helium-core mass, this would
constrain the mass of the remnant and the magnitude of fallback, which are both very important for understanding
the populations of stellar-mass compact objects and the chemical enrichment of galaxies.

All these issues can and need to be addressed quantitatively using high-quality photospheric- and nebular-phase SN II-P
spectra/light-curves, and non-LTE time-dependent radiative-transfer simulations, with allowance for non-thermal excitation/ionisation,
and based on physically-consistent hydrodynamical inputs of SN II-P ejecta.
This is key for providing observational constraints on
the way a generic massive star explodes and the properties of the remnant star left behind.

\appendix

\section{Additional Results}

   In this appendix, we present additional results from our simulations.  These are
   placed here not to divert the main results we wanted to focus on, namely the ejecta kinematics
   and the associated chemical stratification.
   Furthermore, some of these results have already been discussed directly or indirectly
   \citep{falk_arnett_77,baklanov_etal_05,utrobin_07,kasen_woosley_09,woosley_weaver_95,
   WHW02,zhang_etal_08}.

 \begin{figure*}
\epsfig{file=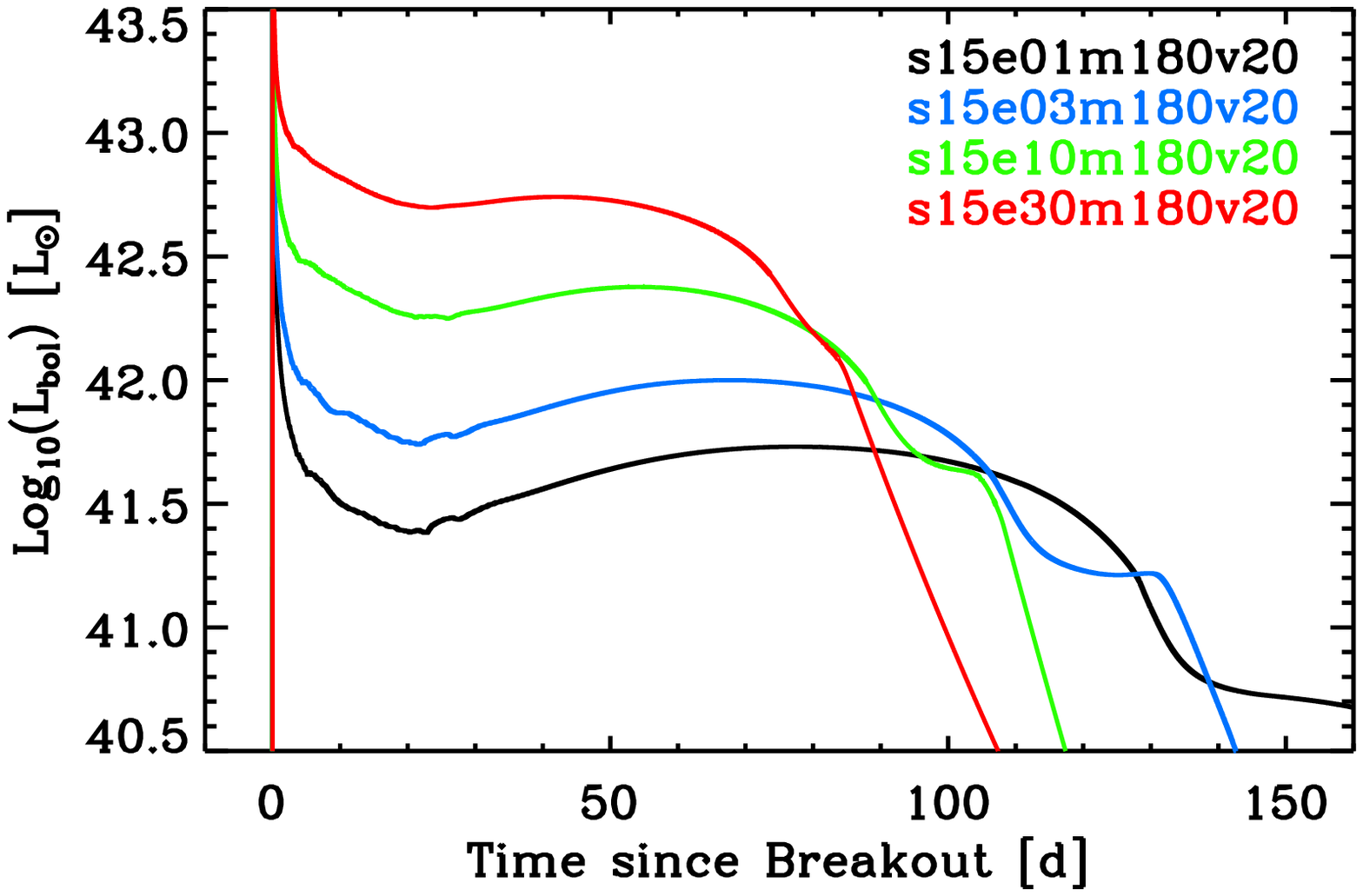,width=8.5cm}
\epsfig{file=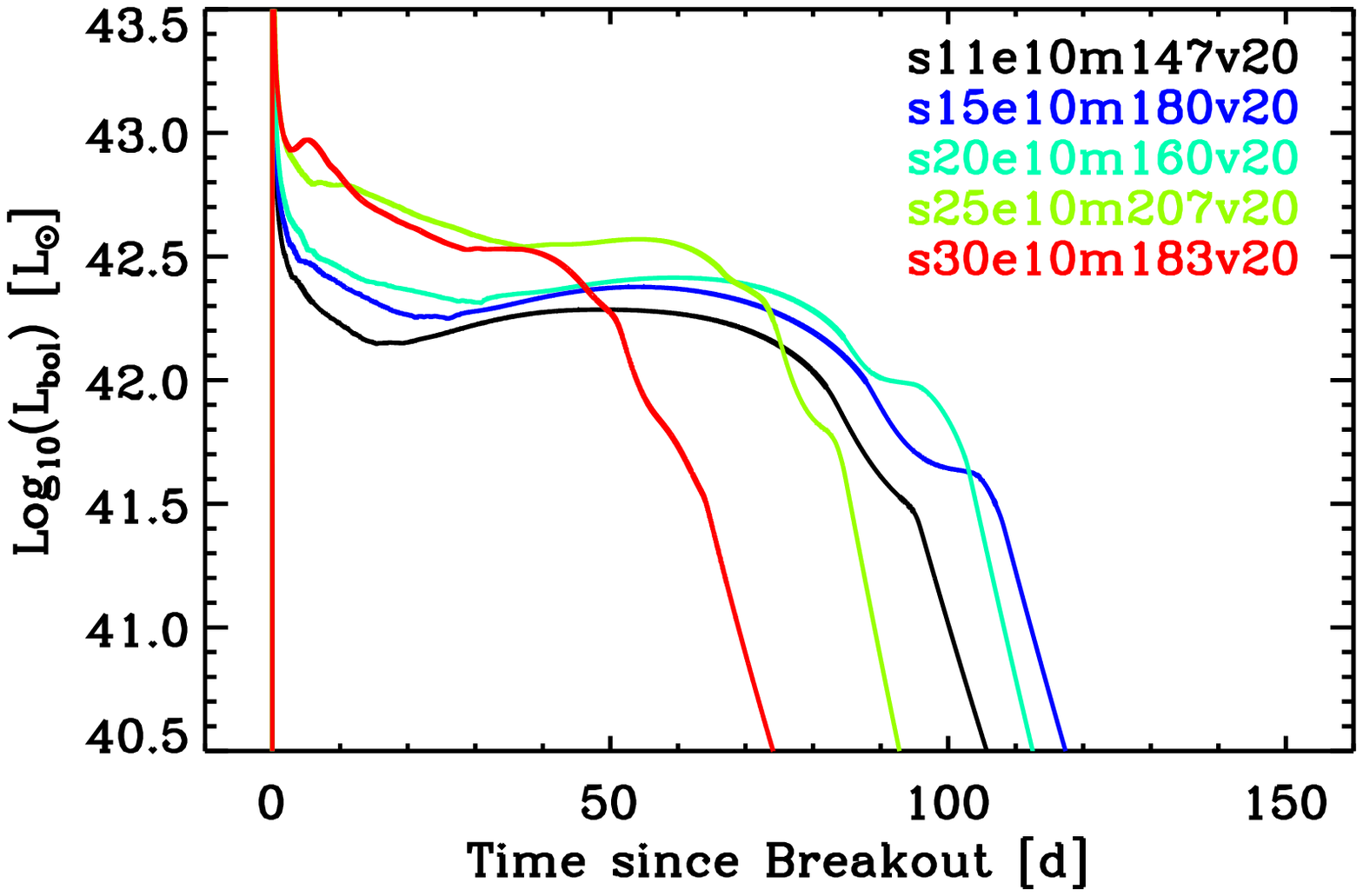,width=8.5cm}
\caption{Evolution of the bolometric luminosity for the first 160 days
for a variety of {\sc v1d} simulations. In the left panel, we show our results
for s15e01m180v20 (black), s15e03m180v20 (blue), s15e10m180v20 (green), and
s15e30m180v20 (red), i.e. simulations that were started from the
WHW02 pre-SN model s15 and exploded to yield and ejecta energy of 0.1, 0.3, 1.0, and 3.0\,B.
In the right panel, we show the result for an ejecta-kinetic energy of 1.0\,B, but now for five different
WHW02 pre-SN models, i.e. s11, s15, s20, s25, and s30.
\label{fig_lbol}}
\end{figure*}

   We give a log of the parameters and the main results of our simulations in Tables~2--6.
For each entry (i.e. simulation name), we first give the simulation parameters, i.e. the initial
model mass on the main sequence ($M_{\rm i}$), the energy deposited by the piston
($E_{\rm kin}$),\footnote{As pointed out earlier, this corresponds to the ejecta kinetic energy
at infinity when the explosion is large, but for low-energy explosions, the large fallback mass
tends to lead to an overestimate of the energy aimed for. The 14th column gives the kinetic energy
the ejecta effectively have asymptotically.}
the Lagrangian mass where the piston is positioned ($M_{\rm piston}$), and the velocity of the piston
($V_{\rm piston}$). We then give the final mass of the pre-SN star ($M_{\rm f}$), the mass of
the compact remnant ($M_{\rm remnant}$, which corresponds to all the material that fails to eject),
and the mass of the ejecta ($M_{\rm ejecta}$, which is just $M_{\rm f}-M_{\rm remnant}$).
Further, we provide information on the ejecta kinematics, with the mass-weighted mean velocity of the
inner 0.1\,\msun\ in the SN ejecta ($V_{\rm inner}$), the velocity at the outer (inner) edge of the oxygen
(hydrogen) shell $V_{\rm e,O}$ ($V_{\rm i,H}$), the photospheric velocity
at 15 and 50\,d after shock breakout ($V_{\rm p,15d}$ and $V_{\rm p,50d}$).
We then give the asymptotic ejecta kinetic energy ($e_{\rm kt}$, which may differ from $E_{\rm kin}$),
the fraction of $e_{\rm kt}$ that resides in the hydrogen-rich envelope ($e_{\rm kfH}$; it is close to 99\% for
low-mass massive stars but decreases for objects with increasing helium-core masses).
Generally, the entire hydrogen-rich and helium-rich shells are ejected in our simulations for any of our
adopted explosion energies and piston characteristics. However, the oxygen-rich shell suffers
dramatically from fallback for low explosion energies and/or piston speeds and thus the
total oxygen yield is given in the table to show this dependence. This correlates directly with the
remnant mass given in the 7th column.
Finally, we give two important times that characterise the light curve. First,
the SN display is born at the shock-breakout time. We thus give here the delay time
between the piston trigger and the shock breakout time $t_{\rm SBO}$.
We also give the plateau duration $\Delta t_P$ (defined as the post-breakout time
when the luminosity has decreased to a tenth of the plateau-peak value).
Note that no $^{56}$Ni is present in our simulations, so that the plateau duration is a lower limit
to what it can be. We show a sample of bolometric-light curves obtained
in our simulations in Fig.~\ref{fig_lbol}, both for a given pre-SN model exploded with various
explosion energies (left panel), and for a range of pre-SN models exploded
with a given explosion energy (right panel). 
Our results for the plateau brightness and duration are in good agreement, for example, 
with the results of \citet{kasen_woosley_09}. We defer a thorough discussion of SNe II-P radiative properties 
to a separate study based on more detailed non-LTE time-dependent simulations,
including both spectra and light curves, and covering the photospheric to the nebular phase
(Dessart \& Hillier, to be submitted).

\begin{figure*}
\epsfig{file=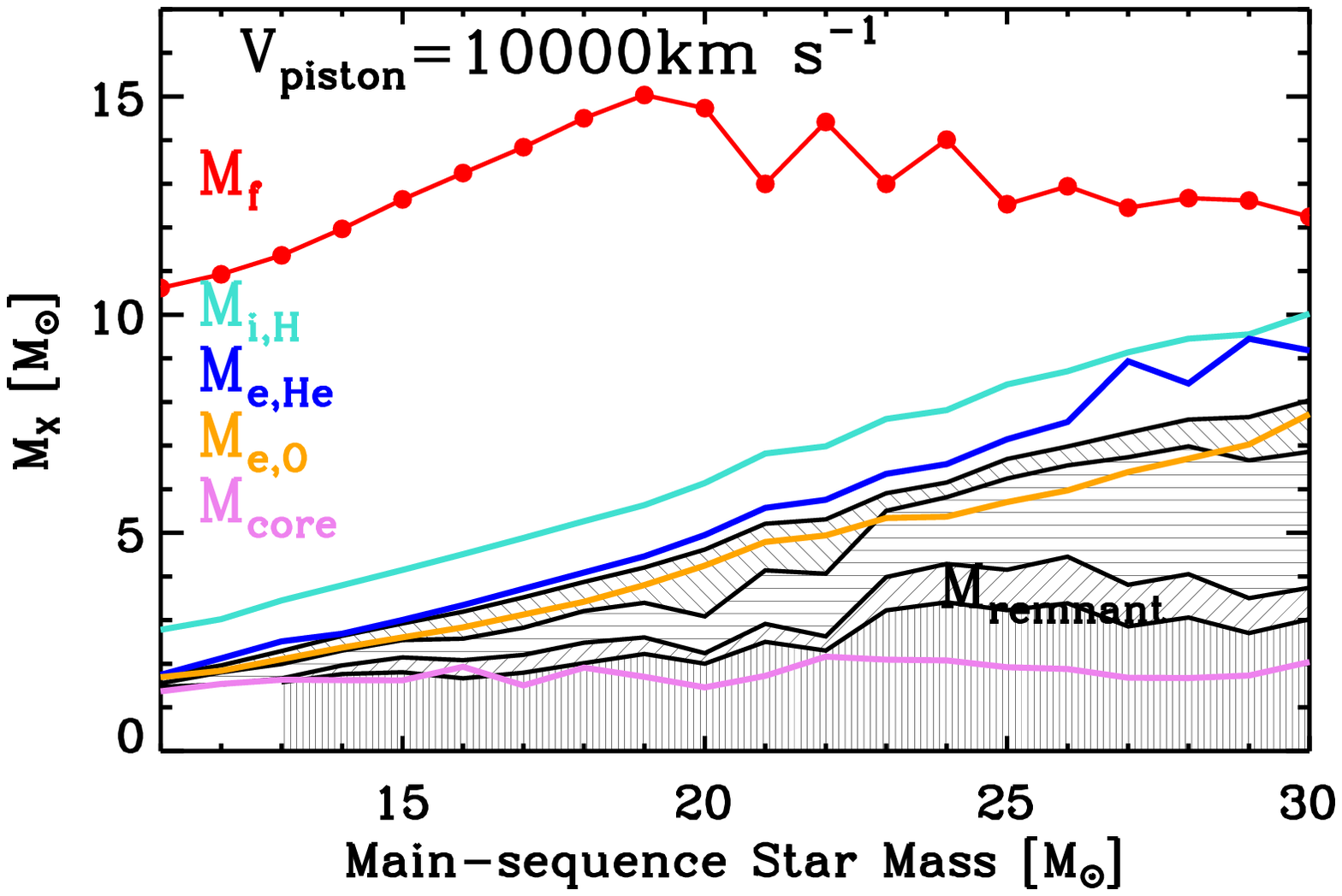,width=8.5cm}
\epsfig{file=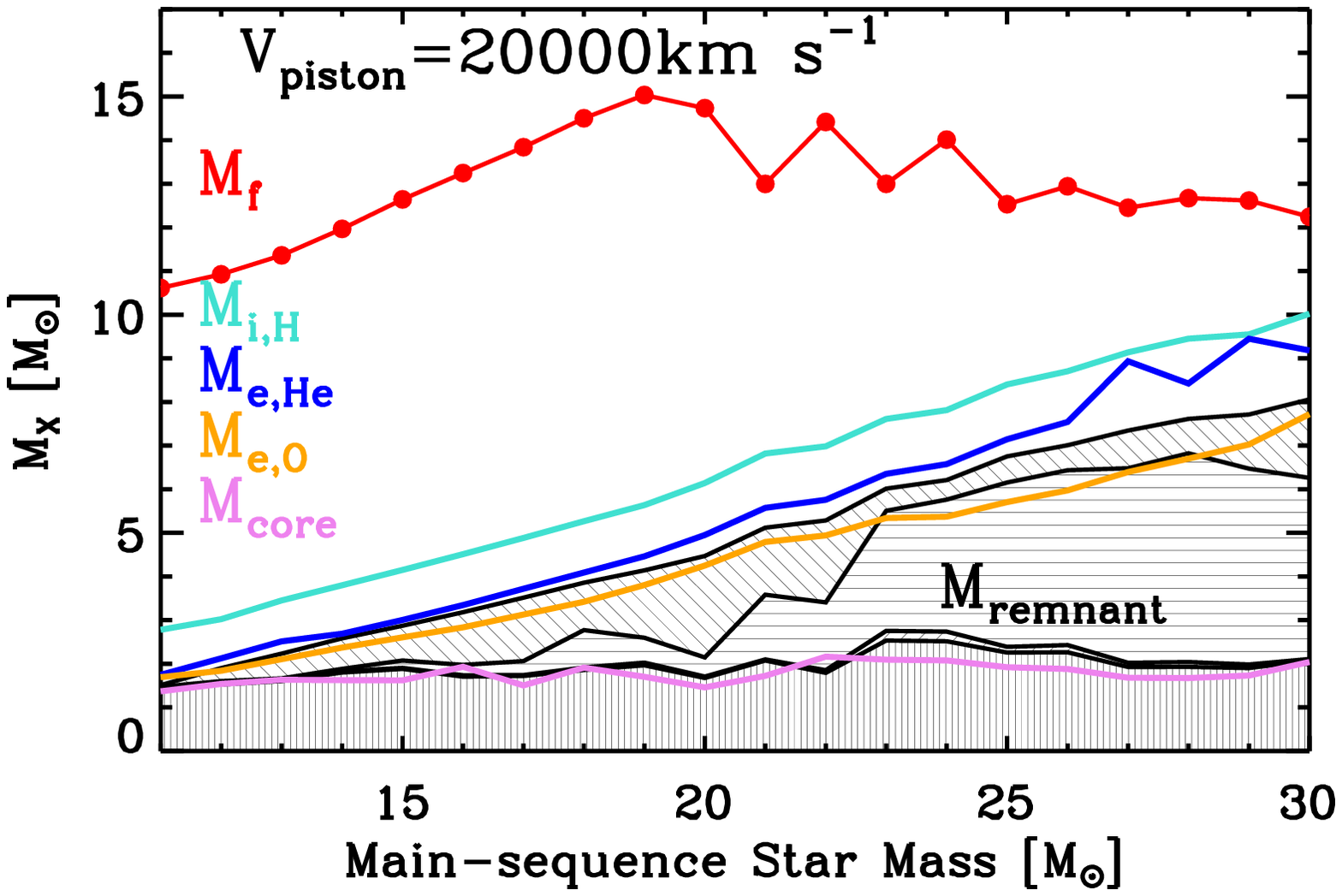,width=8.5cm}
\caption{Illustration as a function of main-sequence mass of several important reference mass coordinates,
characterising both the pre-SN models of WHW02 as well as their resulting ejecta, produced with {\sc v1d}
using a piston speed of 10000\,\kms\ (left) or 20000\,\kms\ (right).
The black curves limiting from above the
hatched regions with stripe orientation at  $-45^{\circ}$, $0^{\circ}$, $45^{\circ}$, and $90^{\circ}$ correspond to models
with asymptotic ejecta kinetic energy of 0.1, 0.3, 1.0, and 3.0\,B (models yield monotonically
increasing remnant masses for decreasing explosion energies).
We over-plot key quantities characterising the structure of the progenitor model (prior to core-collapse and explosion),
such as the final mass (red line), the Lagrangian mass at the base
of the hydrogen shell (turquoise), at the outer edge of the helium/oxygen shell (blue/orange), and the Lagrangian mass
at the iron-core edge (violet).
For ejecta kinetic energy on the order of 0.1\,B, the entire highly-bound helium-core
falls back: Neither nuclear-processed oxygen nor unstable isotopes would be copiously ejected.
Note that the the two simulations based on the pre-SN models s11 and s12, and characterised by an explosion energy of
3.0\,B, and a piston speed of 10000\,\kms, are not plotted since with such a slow piston speed they reach this large explosion
energy on an unphysically long timescale of minutes (they are also omitted from Table~2).
[See text for discussion]. \label{fig_mass_montage}
}
\end{figure*}

    In Fig~\ref{fig_mass_montage}, using all the {\sc v1d} simulations presented above, we show our results
    for important reference masses.
    As a function of main-sequence mass for our sample of pre-SN
    progenitor stars, we show the final star mass at collapse $M_{\rm f}$ (red; dots indicate the initial
    masses of the individual models effectively calculated), the Lagrangian mass coordinate corresponding
    to the inner edge of the hydrogen shell ($M_{\rm i,H}$, turquoise), to the outer edge of the helium core ($M_{\rm e,He}$, blue), to the
    outer edge of the oxygen-rich shell ($M_{\rm e,O}$, orange), and to the outer edge of the iron core ($M_{\rm core}$, violet).
    Notice again the monotonic increase of $M_{\rm i,H}$,  $M_{\rm e,He}$,  $M_{\rm e,O}$, as a function
    of main-sequence mass. Now, over-plotted, we draw black curves (separated by hatched regions of differing
    orientations)  for the remnant mass obtained for each
    pre-SN progenitor model exploded with an energy of 0.1\,B (top black curve), 0.3\,B (black curve second from top),
    1.0\,B (black curve third from top), and 3.0\,B (bottom black curve).
    The piston speed is 10000\,\kms (20000\,\kms) for simulations shown in the left (right) panel.
    Whatever the piston speed, we find that low-energy explosions suffer a large fallback, whose
    magnitude increases with pre-SN model main-sequence mass.
    For example, if all SNe II-P exploded with an energy of 0.1\,B or less,
    no oxygen from the helium core would be ejected, and the helium core would collapse into the neutron star or in the
    black hole. In many cases, despite the successful, albeit somewhat weak, explosion, a large fraction of the core
    would fail to escape.
    For large-energy explosions, our choice of piston speed is not as important and little fallback occurs.
    However, for a standard-energy core-collapse SN explosion of 1.0\,B, the amount of fallback is highly
    dependent on our choice of piston speed, which sets the timescale over which the energy is deposited, as well
    as the strength of the shock. As the shock traverses the helium-core, the slowly decreasing density
    slows it down considerably, so that insufficient energy may be imparted to those deep envelope layers.
    Furthermore, when the SN shock reaches the interface between the hydrogen-rich
    and helium-rich shells, a reverse shock forms and
    slows down these inner regions even further. Together, these two effects tend to lead to larger fallback
    masses for increasing helium-core masses \citep{herant_woosley_94,zhang_etal_08}.
    The dependency on the piston speed is interesting because it constrains the timescale for the explosion.
    Determining observationally the explosion energy, the main-sequence mass and the remnant mass in a given SN
    could thus potentially give a measure of the explosion timescale and constrain the explosion mechanism.

    Using spectroscopic observations and detailed radiative-transfer calculations based on hydrodynamical
    models of SN II-P explosions should allow one to address these issues. The photospheric velocity at 15\,d
    after shock breakout sets a constraint on the explosion energy. The maximum velocity of the oxygen-rich material,
    inferred from nebular spectra using the O{\sc i}\,6303-6363\AA\ doublet line,
    gives a measure of the maximum velocity at which the outer helium core has been ejected, which constrains
    the helium-core mass of the progenitor star (and thus its main-sequence mass).
    Finally, quantitative spectroscopy on the formation of the  O{\sc i}\,6303-6363\AA\ doublet line
    when the ejecta is fully optically thin can determine how much oxygen was effectively ejected. This
    constrains the fraction of the core that was ejected and thus the remnant mass.

\end{document}